\documentclass[useAMS,usenatbib]{mn2e}
\usepackage{etex}
\usepackage{graphicx}
\usepackage{epstopdf}
\usepackage{stfloats}

\usepackage{epsf}
\usepackage{bm}
\usepackage{color}
\usepackage{times}
\usepackage{supertabular}
\usepackage{hyperref}
\usepackage{amsmath}
\usepackage{longtable}
\usepackage{tabularx}
\usepackage{booktabs}
\usepackage{array}
\usepackage{lscape}
\usepackage{amssymb}
\usepackage{threeparttable}
\usepackage{booktabs}
\title[A new  empirical medium-resolution spectral library]{LEMONY--a   Library of Empirical Medium-resolution  spectra
by Observations with the NAOC Xinglong 2.16\,m and YNAO Gaomeigu 2.4\,m telescopes}

\author[Wang et al.] {C. Wang$^{1}$\thanks{E-mails: wchun@pku.edu.cn (CW); x.liu@pku.edu.cn (XWL)},  
                 X.-W. Liu$^{1,2}$\footnotemark[1], Y. Huang$^{1,2}$\thanks{LAMOST Fellow}, 
                 M.-S. Xiang$^{3}$\footnotemark[2], J.-J. Ren$^{3}$, H.-B. Yuan$^{4}$, 
                 \newauthor B.-Q. Chen$^{2}$\footnotemark[2], Z.-J. Tian$^{1}$\footnotemark[2], J.-M. Bai$^{5}$,  N.-C. Sun$^{1}$, F. Zuo$^{3}$, H.-W. Zhang$^{1}$, 
                 \newauthor Y.-W. Zhang$^{5}$,  Z. Fan$^{3}$, A.-L. Luo$^{3}$, J.-R. Shi$^{3}$,  J. Li$^{6}$,  Y.-L. Shao$^{1}$ \\        
$^{1}$Department of Astronomy, Peking University, Beijing 100871, People's Republic of China.\\
$^{2}$South-Western Institute for Astronomy Research, Yunnan University, Kunming, Yunnan 650091, People's Republic of China.\\
$^{3}$National Astronomical Observatories, Chinese Academy of Sciences, Beijing 100012, People's Republic of China.\\
$^{4}$Department of Astronomy, Beijing Normal University, Beijing 100875, People's Republic of China.\\
$^{5}$Yunnan Observatories, Chinese Academy of Sciences, Kunming, Yunnan 650011, People's Republic of China.\\
$^{6}$Department of Space Science and Astronomy, Hebei Normal University, Shijiazhuang 050024, People's Republic of China.}

\begin{document}

\date{20180530}

\pagerange{\pageref{firstpage}--\pageref{lastpage}} \pubyear{except for 2018}

\maketitle

\begin{abstract}

This study expands the coverage and improves the homogeneity of the distribution of MILES template stars in the parameter space, as well as extends the wavelength coverage of the template spectra to the far red beyond the Ca\,{\sc{ii}} triplet. To achieve this  we have carried out a major observational campaign using the OMR long-slit spectrograph mounted on the NAOC 2.16\,m telescope and the YFOSC long-slit spectrograph  mounted on the YNAO 2.4\,m telescope. The original sample is based on the MILES library, supplemented by 918 stars selected from PASTEL database.  In total, 822 OMR and 1,324 YFOSC spectra,  covering respectively the wavelength ranges  $\lambda\lambda$3800--5180 and $\lambda\lambda$5150--9000, have been collected and reduced.  
The spectra have a mean resolution FWHM (full-width at half-maximum) of $\sim 3.3$\,{\AA} and are wavelength- and flux-calibrated to  an accuracy  of $\sim 20$\,km\,s$^{-1}$ and $\sim 5$\,per\,cent, respectively. The spectra are further corrected for systematic errors in the wavelength calibration to an accuracy of $\sim 4$\,km\,s$^{-1}$ by cross-correlating with the theoretical spectra.  Almost all the spectra have an average  signal to noise ratio (SNR) better than 100 per pixel.  Combined with the MILES spectra, there are now   1,731,  1,542, 1,324 and 1,273  stars with  spectra covering respectively  $\lambda\lambda$3800--5180, $\lambda\lambda$3800--7500, $\lambda\lambda$5150--9000 and $\lambda\lambda$3800--9000.  This paper describes  our template star selection, the observation and data reduction, and presents the reduced spectra collected  hitherto.

\end{abstract}


\section{INTRODUCTION}

A low- to medium-resolution spectral library with a good and homogenous  coverage of the
stellar atmospheric parameter space (i.e. effective temperature
$T_{\mathrm{eff}}$, surface gravity $\log\,g$ and metallicity [Fe/H]) is an
important tool in many astronomical applications, from the spectral synthesis analyses of
stellar populations of galaxies  \citep{Guiderdoni1987,Buzzoni1989,Worthey1994,Leitherer1996,1999ApJ...513..224V,Bruzual2003,Leitherer2010,
2012MNRAS.424..157V,Rock2016,Milone2014}  to the stellar atmospheric parameters determinations by
spectral template matching \citep{Adelman2008,Lee20081,Boeche2011,Wuyue,LSP3}. The latter has become increasing important, driven
by a number of already completed or still on-going large scale spectroscopic surveys,
such as the Sloan Extension for Galactic Understanding and Exploration
 \citep[SEGUE;][]{yanny-segue}, the Radial Velocity Experiment \citep[RAVE;
][]{2006AJ....132.1645S}, the Large sky Area Multi-Object fiber Spectroscopic
Telescope (LAMOST) Galactic Spectroscopic Surveys  \citep{2012RAA....12..723Z,
deng-legue, liu-lss-gac, yuan-lamost}, and the APO Galactic Evolution
Experiment  \citep[APOGEE;][]{APOGEE}. These surveys are providing
huge amounts of low- to intermediate-resolution spectral data to help improve our understanding of the structure,
kinematics and chemistry, and formation and evolution of the Milky Way galaxy,
and of other galaxies in general. 

Specifically, the on-going LAMOST Galactic Spectroscopic Surveys have hitherto
collected over 7.5 million low-resolution ($R\sim 1800$) quality optical
spectra  \citep{LAMOST_preface}. 
To derive robust estimates of the stellar atmospheric parameters as
well as the  radial velocities from this huge data set, two stellar parameter
determination pipelines based on the technique of spectral template matching, the LAMOST Stellar Parameter Pipeline  \citep[LASP;
][]{lamost-dr1}  and the LAMOST Stellar Parameter Pipeline at Peking University
 \citep[LSP3; ][]{LSP3}, have been developed. LASP determines both the atmospheric parameters as
well as the radial velocities by template matching with the empirical spectral
library ELODIE  \citep{2001A&A...369.1048P}, whereas LSP3 uses ELODIE for the radial
velocity determinations and another empirical spectral library MILES
 \citep{2006MNRAS.371..703S,2011A&A...532A..95F} for the atmospheric parameter
determinations.
 
Several either empirical \citep{2006MNRAS.371..703S,2011A&A...532A..95F,STELIB,ngsl,xsl,indo-us,2012MNRAS.424..157V,2001MNRAS.326..959C,Jones1999} or  synthetic \citep{Lastennet2002,Barbuy2003,Murphy2004,Munari2005,Martins2005,Coelho2005} spectral libraries are now available (cf.
the recent review by \citealt{Wuyue}). Compared to synthetic spectra, empirical
spectra have the advantage that they represent {\em real} stars. On the other
hand, empirical spectral libraries are always restricted by the available
observations and thus limited in term of the parameter space coverage \citep{Martins2007,Wuyue}. 
Either due to the lack of a wide coverage of the  stellar atmospheric parameter space or insufficiently accurate  flux-calibration over a wide wavelength range,
 most of those currently available  empirical spectral libraries are not suitable as template spectral libraries  
  for the determinations of  the  stellar atmospheric parameters. For examples, STELIB  \citep{STELIB}, NGSL  \citep{ngsl} and XSL  \citep{xsl} have insufficient atmospheric parameter space coverage, while  INDO-US  \citep{indo-us} and MIUSCAT  \citep{2012MNRAS.424..157V} suffer from poor flux-calibration, and  CaT  \citep{2001MNRAS.326..959C} and Jones  \citep{Jones1999}  have narrow wavelength coverages. 
In comparison, MILES and ELODIE, adopted as the template library  respectively LASP and LSP3,  have  broad atmospheric parameter  coverages as well as   wide wavelength ranges.   MILES spectra were observed using a long-slit spectrograph with a
spectral resolution comparable to that of the LAMOST spectra and were
accurately flux-calibrated to a few per cent  \citep{2006MNRAS.371..703S}. In comparison,  ELODIE  spectra were obtained by the ELODIE spectrograph with medium ($R\sim$10,000) to high spectral resolution ($R\sim$42,000)  and were poorly flux-calibrated. Thus
the MILES spectral library is ideal for the determinations of the atmospheric parameters, although
ELODIE works much better for the radial velocity determinations, given the much
higher spectral resolution of the spectra of the latter  \citep{LSP3}. With MILES, LSP3 has
achieved a precision of 150 K, 0.25 dex, 0.15 dex for the determinations of
$T_{\mathrm{eff}}$, log\, $g$, and [Fe/H], respectively, for LAMOST spectra of
FGK type stars of signal-to-noise ratios (SNRs) per pixel better than 10. 

As discussed in \cite{LSP3} and \cite{yuan-lamost}, there are several aspects
of the MILES library that are desired of further improvement. Firstly, while
the library has a decent coverage of the stellar parameter space, the
distribution of stars in the parameter space is not uniform -- there are
clusters and holes of stars in the distribution that produce systematic
errors in the derived atmospheric stellar parameters  \citep{LSP3}.  A more
acute problem is that the spectra only cover  wavelength ranges up to
7500\,{\AA}. Given that the LAMOST Galactic Spectroscopic Surveys target stars
of all colors in the Galactic disk and halo, there are a significant fraction
(about 30\,per\,cent) of spectra that meet the survey SNR requirement, i.e. better
than 10, in the red, but fail to do so in the blue, either because the stars
are intrinsically red (i.e. of late spectral types) or heavily reddened by the
interstellar dust grains, or both. The lack of suitable template spectra in the
red is largely responsible for the fact that of all the spectra hitherto
collected by LAMOST, only just over half have the atmospheric parameters
determined  \citep{lamost-dr1, yuan-lamost}. Although there are several empirical  libraries e.g.\,
MIUSCAT, INDO-US, STELIB, NGSL and  XSL as mentioned above,  that provide spectra covering almost
the whole optical wavelength range, those spectra
are  either poorly flux-calibrated or have insufficient coverage of the atmospheric parameter space. Thus it is highly desirable to extend the
wavelength coverage of all MILES template spectra to the far red, beyond the
Ca\,{\sc ii} triplet for example.  Finally, there is also room of expansion in the 
parameter coverage of MILES template stars, particularly toward low metallicities and low effective 
temperatures. With the above considerations in mind, we
have embarked on a massive campaign to expand the MILES empirical spectral
library. To maintain the maximum internal consistency, all spectra will be
collected using long-slit spectrographs with a spectral resolution comparable
to that of the MILES spectra and accurately flux-calibrated to a few
per cent. Additional template stars with accurately known atmospheric
parameters, mostly determined with high resolution spectroscopy, are selected
and added to the library, and observed in order to increase the parameter
coverage as well as to improve the homogeneity of the distribution of template stars in
the parameter space. Spectra extending to the far red will also be collected such that
LAMOST spectra of stars of either intrinsically red colors or heavily reddened by dust grains
can also be properly analyzed with LSP3. The spectra collected will also be of
interest for other  applications such as  the spectral synthesis analyses of stellar
populations of galaxies as mentioned above.

The  observed  high SNR (better than 100 per\,pixel) spectra  collected in  LEMONY--a Library of Empirical Medium-resolution spectra by Observations with the NAOC Xinglong 2.16 m and YNAO Gaomeigu 2.4 m telescopes,  are  
accurately  wavelength-  and flux-calibrated,  and cover almost the whole optical
range of $\lambda \lambda $ 3800--9000 at a FWHM (full-width at half-maximum)
resolution of $\sim 3.3$\,{\AA}. The spectra were  collected with the NAOC
2.16m telescope in the blue and the YNAO 2.4m in the red.

Here we present the final results of this observational campaign.  
The  paper is organized  as follows.  In Section\,2  we describe the selection of
additional template stars and  observations.  
The data reduction is presented  in Section\,3.  We describe the new library LEMONY to guide its use in Section\,4. 
Qualities of the secured spectra are examined and discussed in Section\,5. 
In Section\,6, we discuss some  improvements of LSP3 based on  LEMONY. 
Finally, Section\,7 summarizes the main results of the paper.

\section{OBSERVATIONS}
\subsection{Target selection}

As described above, the observational campaign to provide a new stellar library has two goals.  The first is to increase the coverage in comparison to MILES library and  improve the homogeneity of
the distribution of the template stars in the parameter space. For this purpose, we have
selected 918 additional template stars from the PASTEL
 \citep{2010A&A...515A.111S} catalogue, a bibliographical compilation of accurate
stellar atmospheric parameters including $T_{\mathrm{eff}}$, $\log\,g$ and
[Fe/H], mostly determined with high resolution, high signal-to-noise
spectroscopy. By 2013, the catalogue  contains 18,096 atmospheric parameter
measurements for 8,428 unique stars, collected from 865 bibliographical
references. For a given star, if one of the parameters has more than one
determinations, the average is adopted for that parameter after $3\sigma$-clipping. 
Adopting the simple averages of measurements  available in the
literature may however lead to some systematic errors as the values collected
from different sources do not always have the same quality and are often
determined using different techniques. Also note that not all values of
effective temperature and surface gravity listed in PASTEL were determined with
high resolution spectroscopy. To solve this problem, we have recalibrated the
measurements of stellar atmospheric parameters collected in PASTEL (that
includes essentially all stars in the MILES library) from a
variety of sources such that they are on the same scale. The effective temperatures are recalibrated using  the empirical 
metallicity-dependent calibrations of effective temperature against colours presented by \cite{Huang2015}.  The surface gravities are directly derived from the stellar radii and masses. 
The stellar radii are  determined using the Hipparcos \citep{Perryman1997} distances  and   the effective temperatures.  The stellar masses are determined from the stellar evolution models based on the measured stellar luminosities, effective temperatures and metallicities.  The metallicities are recalibrated using 34 benchmark stars of different spectral types, with measured metallicities accurate to  better than 0.02\,dex \citep{Jofre2015}.  Considering that the determination  of one   atmospheric parameter  depends on the values of  the other parameter, we iterate the  whole recalibration process until all the measurements  converge.
The results will be presented in a separate paper (Huang et al. 2018, in preparation). 

\begin{figure}
\centering
\includegraphics[width=3.5in]{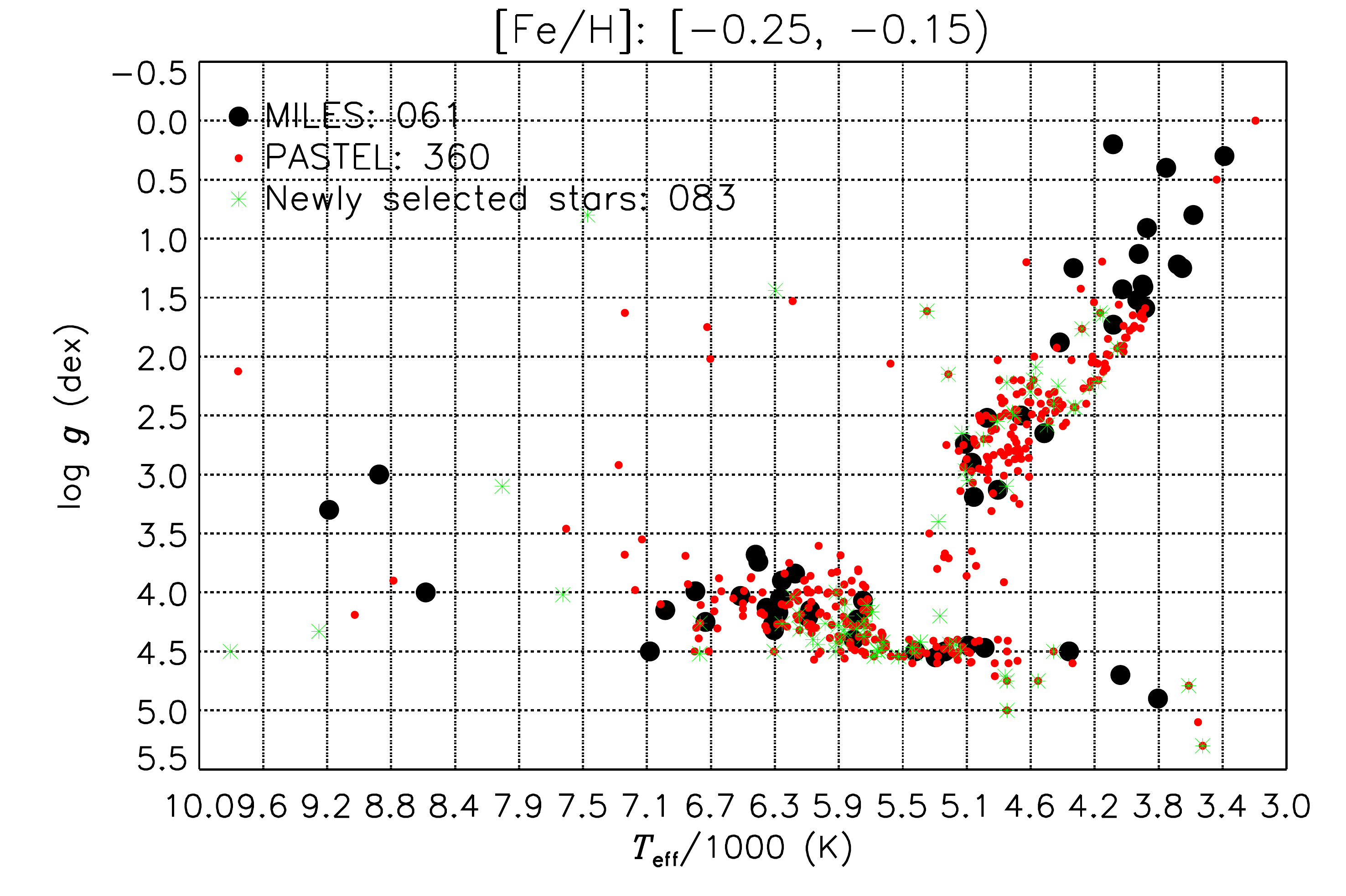}
\caption{Selecting new template stars for the metallicity bin $-0.25 \leq 
{\rm [Fe/H]} < -0.15$\,dex. Stars from the MILES library, from the PASTEL 
catalogue and those newly selected ones are marked by large black dots, small 
red dots and small green starry dots, respectively.}
\label{figure1}
\end{figure}
\begin{figure*}
\centering
\includegraphics[width=5.0in]{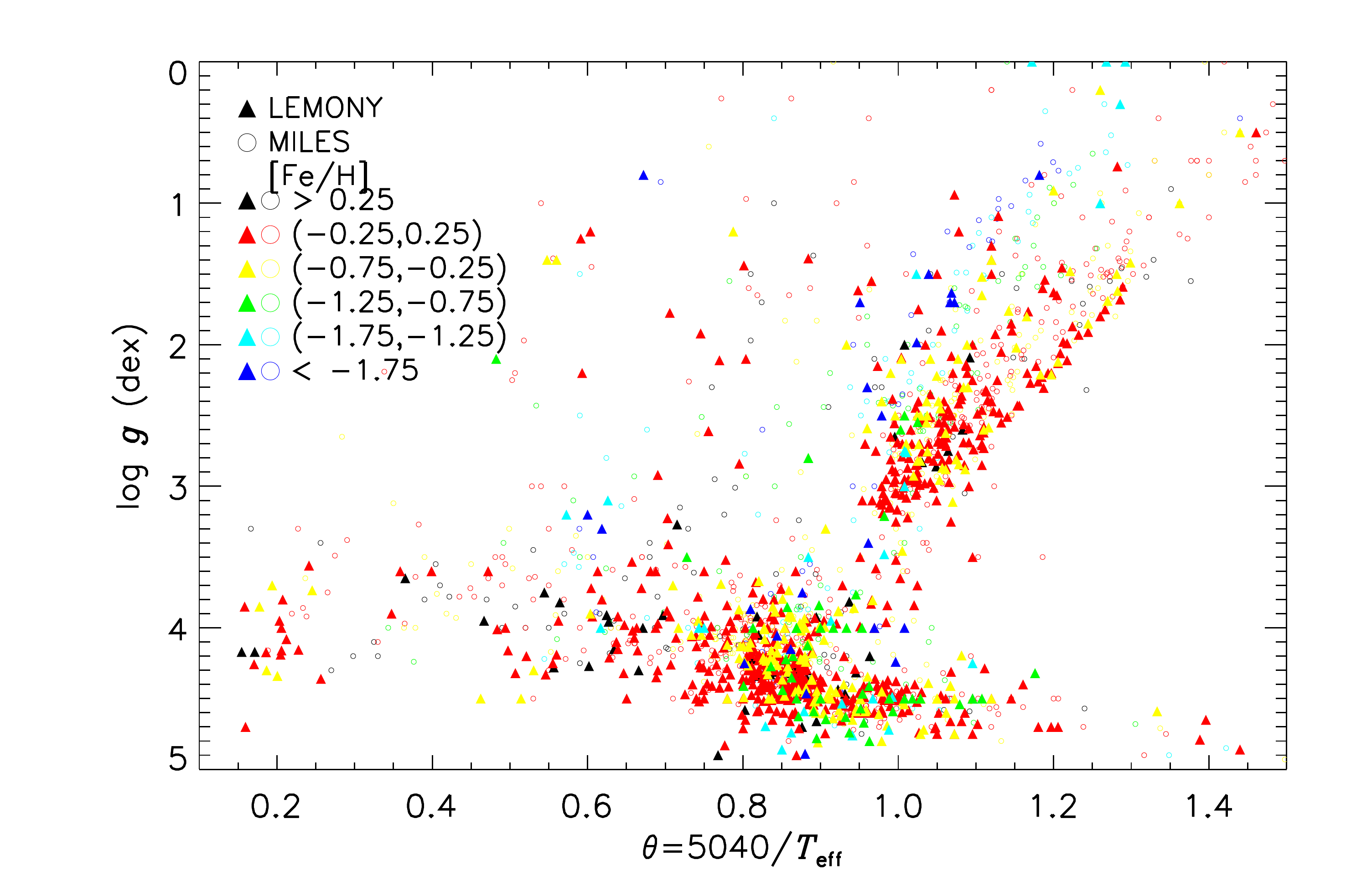}
\caption{Distribution of template stars in the LEMONY library 
in the $T_{\rm eff}$--$\log\,g$ plane. Different symbols are used for stars 
from the MILES library and for newly added ones, whereas different colors 
are used to indicate stars in different metallicity bins, as marked in the top
left corner of the diagram.}
\label{figure2}
\end{figure*}
\begin{figure*}
\centering
\includegraphics[width=5.5in]{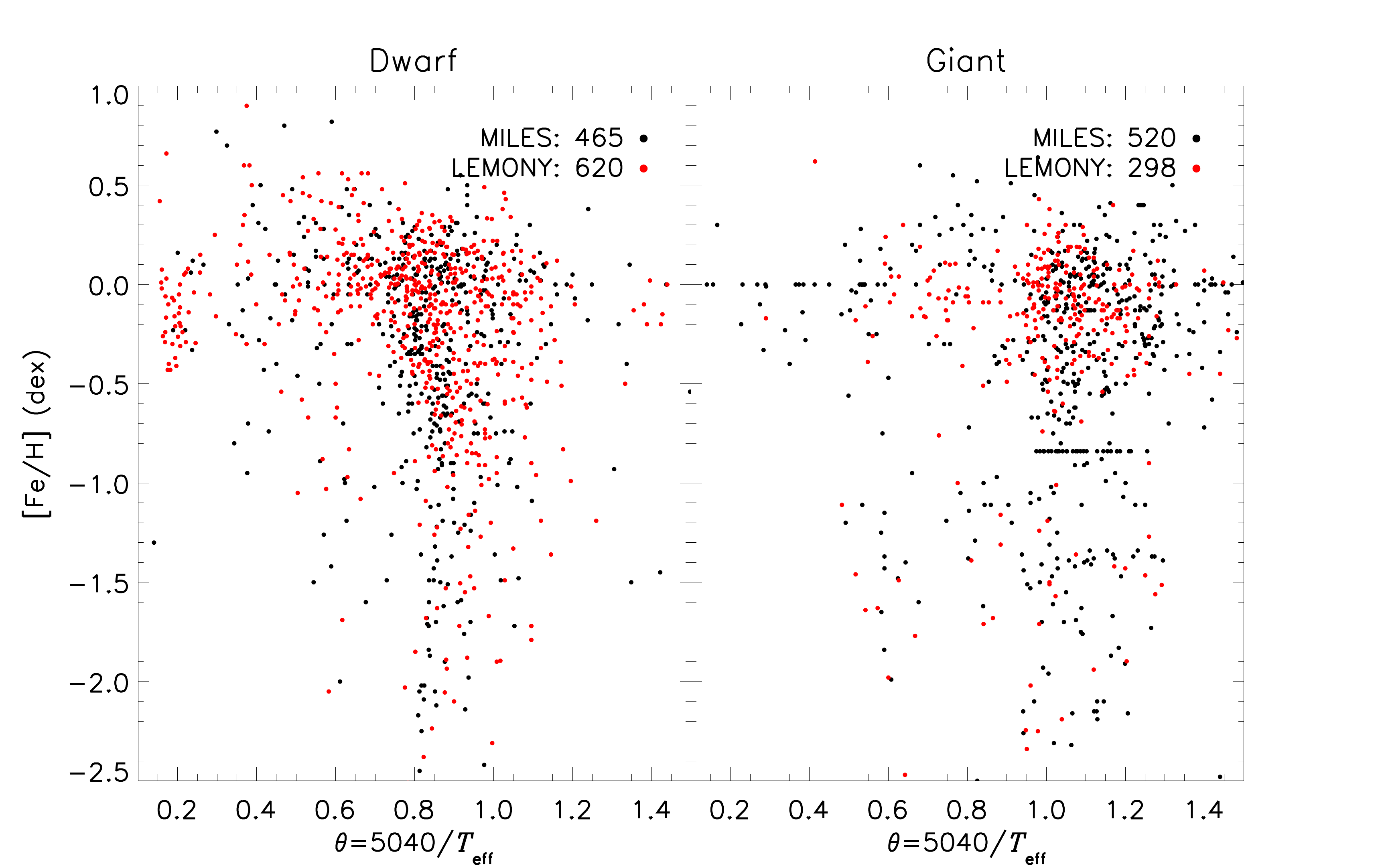}
\caption{Same as Fig.\,2 but in the $T_{\rm eff}$--[Fe/H] plane. Also dwarfs 
($\log\,g > 3.5$\,dex; left panel) and giants ($\log\,g \le 3.5$\,dex; right
panel) are plotted separately. Black and red dots denote stars 
from the MILES library and the newly added ones, respectively.}
\label{figure3}
\end{figure*}

The newly selected  918 template stars consist of two parts. Firstly, we select
all stars in PASTEL of effective temperatures higher than 10,000\,K or lower
than 3,000 K considering the extreme scarcity of such stars in the  MILES
library. Secondly, all stars of $T_{\mathrm{eff}}$ between 3,000 and 10,000\,K
in the MILES library and in the PASTEL catalogue are grouped into
several metallicity bins, as listed in Table~\ref{table1}. Then stars in a
given metallicity bin are plotted on a $T_{\mathrm{eff}}$--$\log\,g$ diagram
with grids of steps of 200\,K and 0.25\,dex in $T_{\mathrm{eff}}$ and
$\log\,g$, respectively. If available, new template stars are then selected
from the PASTEL catalogue to ensure that the LEMONY library contains at least
one template star in a given grid cell. As an example, Fig.\,\ref{figure1}
plots stars from the MILES library and from the PASTEL catalogue in the
$T_{\mathrm{eff}}$--$\log\,g$ plane for the metallicity bin $-0.25 \leq {\rm
[Fe/H]} < -0.15$\,dex. The new template stars selected from the PASTEL catalogue are also
marked. The numbers of newly selected stars and those in the MILES
library for the individual metallicity bins are listed in Table~\ref{table1}.
The distribution  of all template stars in the LEMONY library in the
$T_{\mathrm{eff}}$--$\log\,g$ and $T_{\mathrm{eff}}$--[Fe/H] planes are
presented in Figs.\,\ref{figure2} and \ref{figure3}, respectively. These figures
show that the LEMONY library  has a broader parameter
space coverage than the MILES and some of the holes apparent in the
distribution of the MILES stars are now filled up by the newly
selected targets.


\begin{table*} 
\centering
\caption{Metallicity bins used to select new template stars, and
the numbers of the MILES and the newly added template stars in the individual bins.}
\begin{tabular}{cccccc}
\hline
[Fe/H] metallicity bin (dex)&[$-$5.0,$-$1.75)&[$-$1.75,$-$1.25)&[$-$1.25,$-$1.0)&[$-$1.0,$-$0.75)&[$-$0.75,$-$0.65) \\
Number of stars in the MILES & 44 & 71 & 35 & 64 & 34\\
Number of  newly added stars & 33 & 37 & 20 & 37 & 16 \\
\hline
[Fe/H] metallicity bin (dex)&[$-$0.65,$-$0.55)&[$-$0.55,$-$0.45)  &[$-$0.45,$-$0.35)&[$-$0.35,$-$0.25)&[$-$0.25,$-$0.15) \\
Number of stars in the MILES & 33 & 44  & 43& 75 & 66 \\
Number of  newly added stars & 23 & 33 & 61&  50 & 83 \\
\hline
[Fe/H] metallicity bin (dex)&[$-$0.15,$-$0.05)&[$-$0.05,0.05)&[0.05,0.15)&[0.15,0.25)&[0.25,1] \\
Number of stars in the MILES & 72 & 105 & 104 & 46 & 79 \\ 
Number of newly added stars  & 126 & 148 &  103 &  76 &72 \\
\hline 
\end{tabular}
\label{table1}
\end{table*}

\subsection{Observations}

To extend the wavelength coverage of the  template spectra to the far red, the YFOSC long-slit spectrograph mounted on the YNAO 2.4\,m telescope is used with a G8 grism to cover the wavelength range 5150--9800\,{\AA}. An EEV
2048$\times$4068 CCD of 13.5$\mu$m pixel size was used as detector. The G8
grism has a dispersion of 90\,{\AA}\,mm$^{-1}$, yielding 1.22\,{\AA} per pixel
at the detector. This setup was used to collect red spectra of all stars in the LEMONY library.  All stars were observed with the YFOSC spectrograph with a slit width of 0.58\,arcsec projected on the sky, yielding a FWHM resolution of $\sim
3.6$\,{\AA}.  Right after the science exposures, exposures of He-Ne lamp were obtained to wavelength-calibrate the YFOSC spectra. 

The blue spectra of the newly added stars are collected with the OMR long-slit spectrograph mounted on the NAOC 2.16\,m telescope using a 1200B grating,  covering 3800--5180\,{\AA}. A SPEC10 1340$\times$400 CCD of 20$\mu$m pixel size was used as detector. The 1200B grating has a dispersion of 50\,{\AA}\,mm$^{-1}$, yielding 1.03\,{\AA} per pixel at the detector.  A slit width of 1.8\,arcsec was used  to obtain the blue spectra with a FWHM resolution of $\sim 2.8$\,{\AA}. The exposures of He-Ar lamp were also collected to wavelength-calibrate the OMR spectra. 

For the purpose of accurate flux calibration, all stars were also observed with a wide slit of widths 5.05 and 9.5 \,arcsec with the YFOSC and OMR spectrographs, respectively.  Those same wide slits were used to observe
spectral standard stars for the flux-calibration. After flux-calibration, the
spectrum of a template star obtained with the narrow slit was scaled with a
low-order polynomial function to match the spectral energy distribution (SED)
of the same star obtained with the wide slit.


\begin{figure*}
\centering
\includegraphics[width=7.0in]{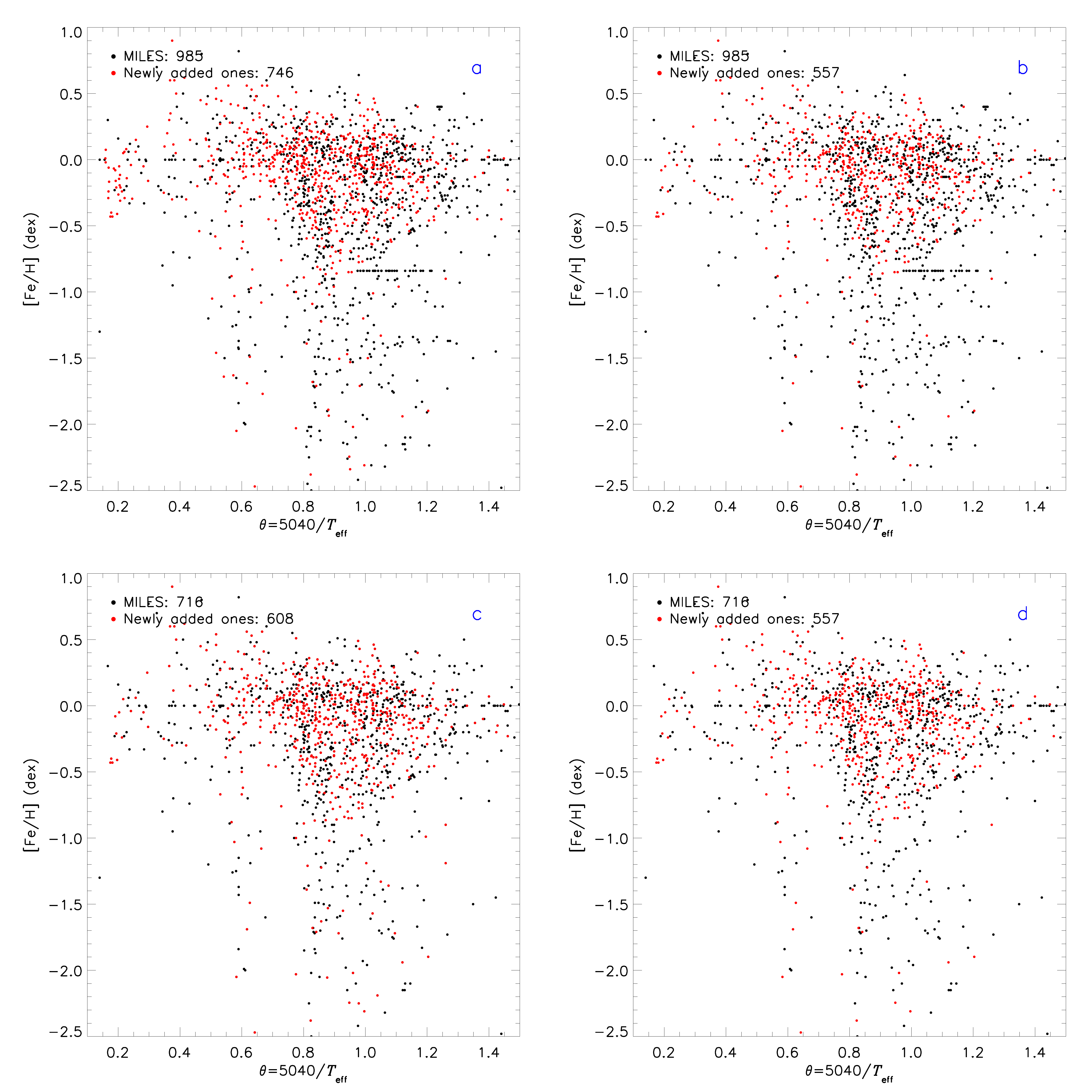}
\caption{The distributions of stars in the LEMONY library  with   spectra of  different wavelength coverages in the $T_{\rm eff}$--[Fe/H] plane. Panels a,b,c,d shows  stars with spectra covering of 3800-5180, 3800-7200, 5150-9000 and  3800-9000\,\AA, respectively.  Black and red dots denote stars  from the MILES library and the newly added ones selected  from the PASTEL catalogue, respectively.}
\label{spectra_distribution}
\end{figure*}

During the 2013--2017 observational season, 1,324 red spectra were collected in 
134 (partial) observing nights with the YNAO 2.4\,m telescope and 822 blue spectra 
were obtained in another 58 nights with the NAOC 2.16\,m telescope.   
LEMONY contains 822 blue spectra and 1,324 red spectra. Including the spectra of  MILES,  one now has 1,731 stars with
 spectra   covering   3800-5180\,\AA, 1,542 stars   with spectra  covering  3800-7500\,\AA,  1,324 stars with spectra 
  covering 5150-9000\,\AA, and 1,273 stars with spectra covering  3800-9000\,\AA.  Fig.\,\ref{spectra_distribution} plots the distributions of 
 those four groups of   stars   in the  $T_{\rm eff}$--[Fe/H] plane. The Figure shows that template
  stars with spectra covering  3800-7500\,\AA \,cover a broader parameter space than the MILES stars and some of the 
  holes apparent in the distribution of the  MILES stars are now filled up by the 
 newly observed targets.  The  1,324 stars with YFOSC red spectra also have a broad 
 parameter space coverage. These stars can be used as templates
 to estimate the atmospheric parameters from the LAMOST red-arm spectra for stars that are either 
 intrinsically red (i.e. of late spectral types) or heavily reddened by the interstellar dust grains. 
Compare to ELODIE, the 1,542 stars   with spectra  covering  3800-7500\,\AA \, contain more metal-poor giant stars.

\section{data reduction}
All LEMONY spectra were reduced  with IRAF\footnote{\begin{scriptsize}IRAF
\end{scriptsize} is distributed by the National Optical Astronomy
Observatories, USA, which are operated by the Association of Universities for
Research in Astronomy, Inc., under cooperative agreement with the National
Science Foundation, USA.}. The reduction included several steps: bias
subtraction, flat-fielding, cosmic ray cleaning, sky background subtraction, 1D
spectrum extraction, wavelength calibration and (relative) flux calibration.
The wavelength calibration and relative flux calibration are the essential
steps affecting the spectral quality. In the following subsections, we will
describe these two steps in some detail and the accuracy achieved.

\subsection{Spectral SNR and resolution}

\begin{figure}
  \centering
  \includegraphics[width=3.5in]{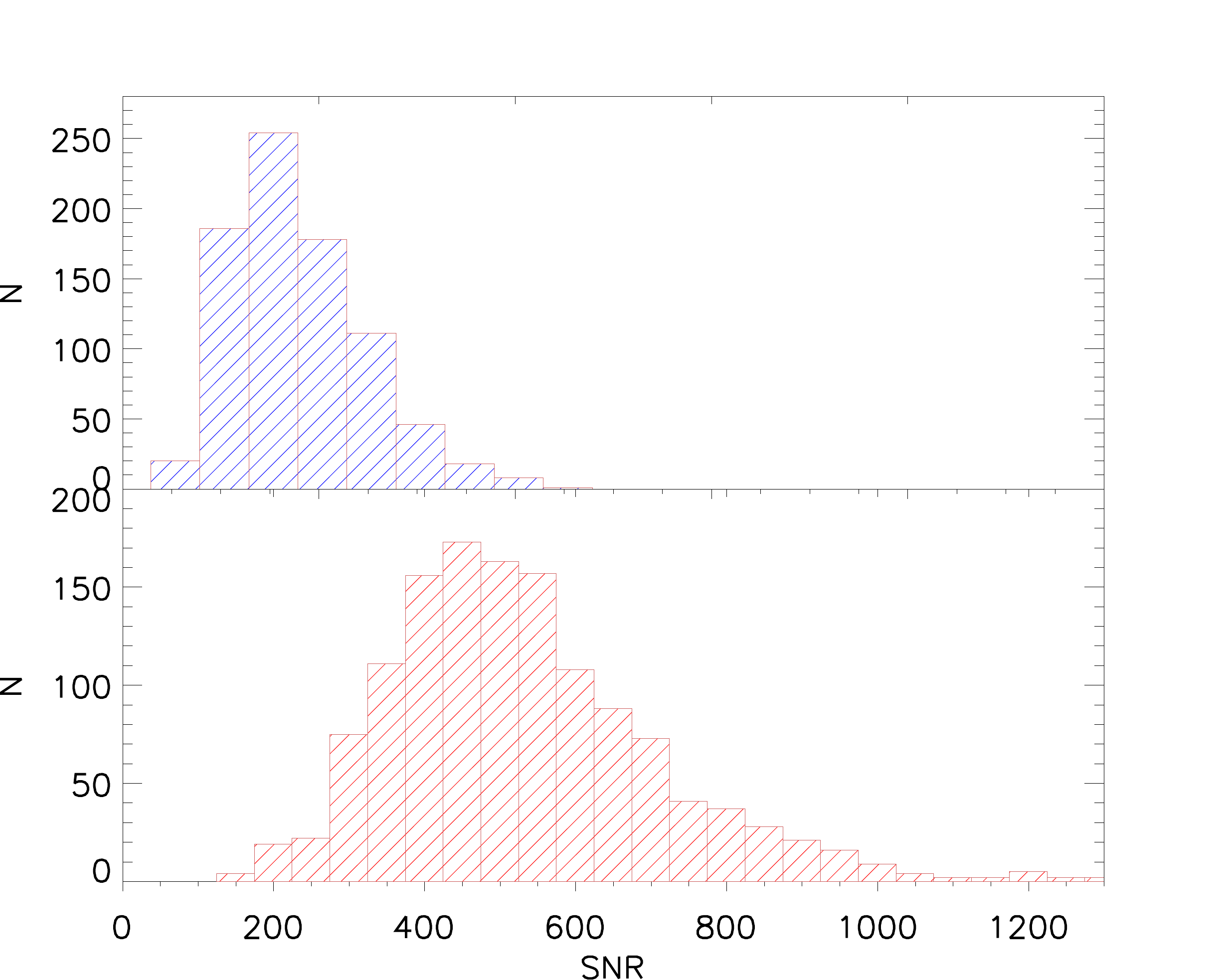}
  \caption{Distributions of mean SNRs of  the LEMONY blue (top) and red 
  (bottom) template spectra.}
  \label{figure4}
  \end{figure}

For the purpose of being used as the template spectra for the derivation of stellar
atmospheric parameters from LAMOST spectra, the template spectra 
should have good SNRs and a resolution comparable to that of LAMOST spectra.
The mean SNRs over the whole wavelength range of  almost all  our LEMONY 
spectra are  higher than 100, except for 81 blue spectra,
but still essentially  all  higher than 50.  The stars with blue spectra of SNRs lower 100 have  very low  effective  temperatures,
 leading to few photons in blue parts of the spectra. Fig.\,\ref{figure4} plots the SNR distributions of
the  LEMONY blue and red template spectra. 
  
\begin{figure}
\centering
\includegraphics[width=3.5in]{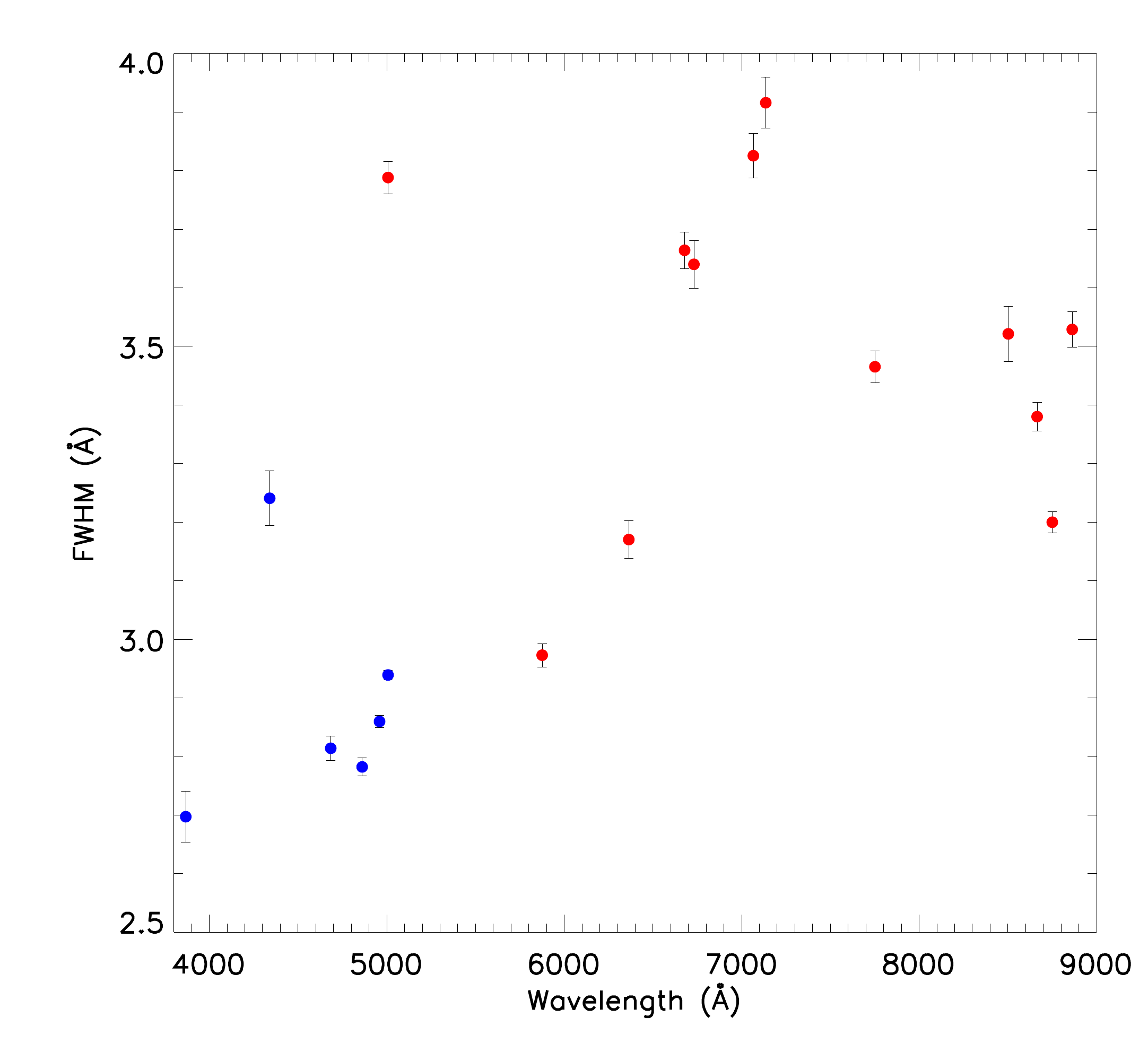}
\caption{FWHMs of emission lines derived from the blue (blue dots) and red (red dots) spectra of 
planetary nebulae  NGC\,2392  and IC\,4997 observed with the same instrumental setups 
used for our template stars. }
\label{figure5}
\end{figure}

To estimate the spectral resolution as a function of wavelength, we make use of the
spectra of planetary nebulae IC\,4997 and NGC\,2392 obtained with  same  instruments.
The nebulae have  rich emission lines across the whole optical spectral wavelength range  and expansion velocities much
smaller than our spectral resolution. The results are plotted in
Fig.\,\ref{figure5}. As the Figure shows, the template spectra have an average
resolution FWHM of approximately 2.8\,{\AA}, 3.6\,{\AA} and 3.3\,{\AA} for the blue, red and
the whole spectra, respectively.


\begin{figure*}
\centering
\includegraphics[width=7.0in]{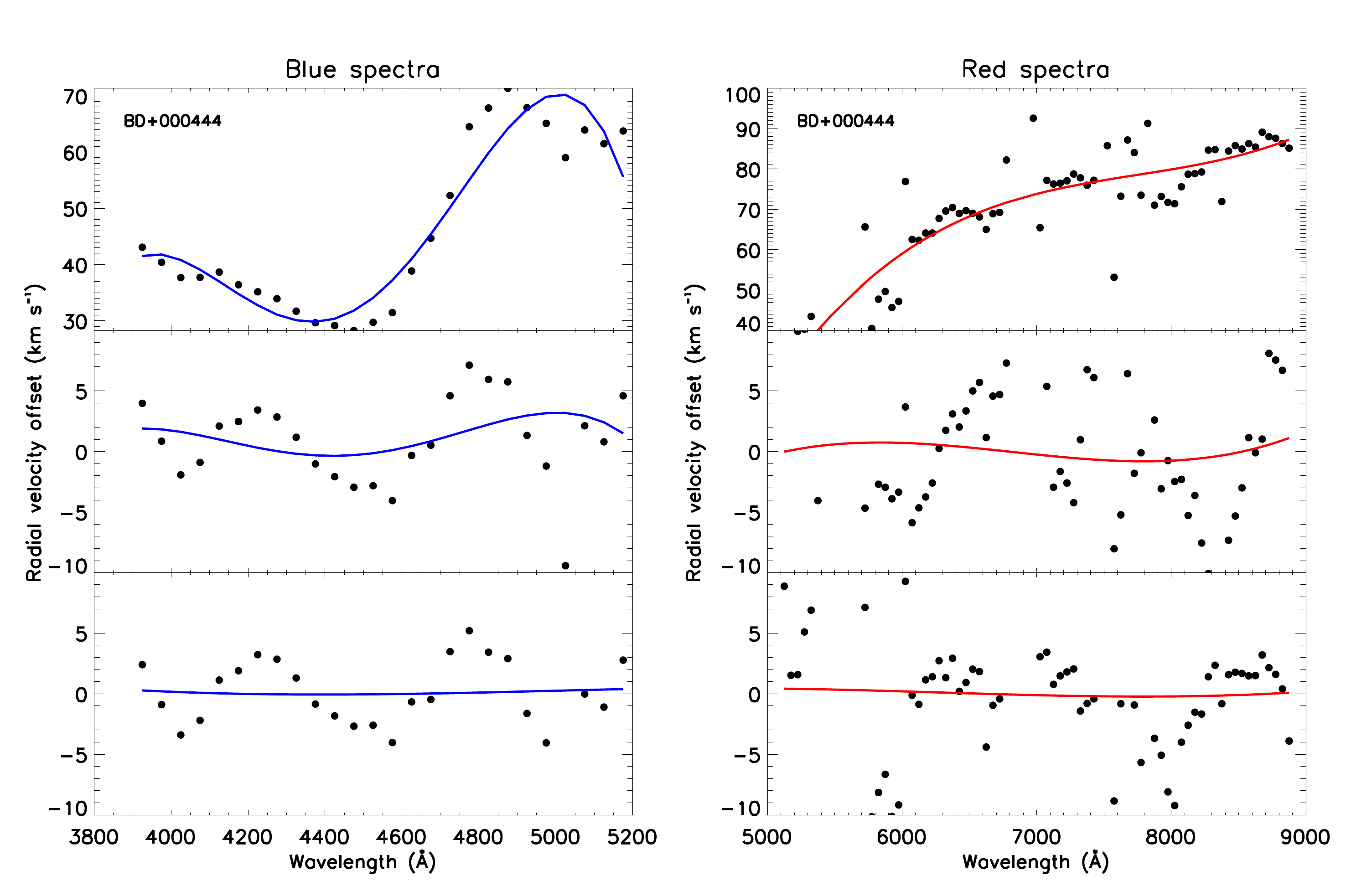}
\caption{The radial velocity offsets as a function of wavelength for the blue (left panels) and red spectra (right panels) of BD+000444. The top, middle and bottom panels show the results before the corrections, and after the first and second round of   corrections, respectively.}
\label{plot_rv}
\end{figure*}

\begin{figure}
\centering
\includegraphics[width=3.5in]{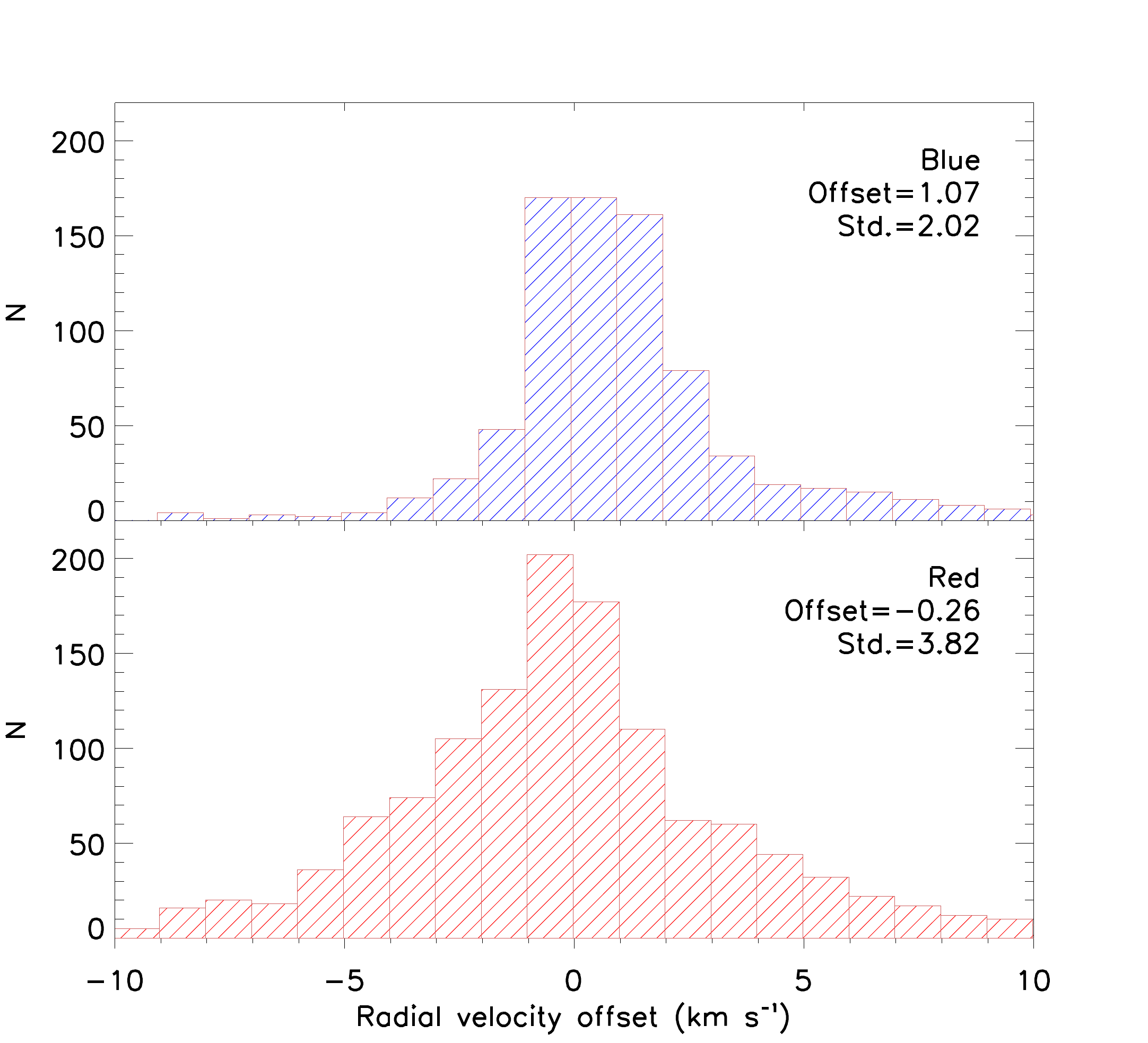}
\caption{Histograms of radial velocity offsets yielded by the blue (top panel)
and red (bottom panel) spectra after corrected for the systematic errors in the 
spectral dispersion relation deduced from the arc spectrum.}
\label{figurew}
\end{figure}

\subsection{Wavelength calibration}

As mentioned above, the spectra secured with the OMR and YFOSC spectrographs
were respectively wavelength-calibrated using He-Ar and He-Ne arc exposures
taken after the target observation.  The root mean square (RMS)
residuals of wavelength-calibration were around 0.1\,{\AA}.  The
spectra were then converted to heliocentric wavelengths after corrected for the
diurnal and annual motions of the Earth.  By comparing the radial velocities
derived from our spectra with LSP3 with those of \cite{G06} (with a mean
accuracy of 0.7\,km\,s$^{-1}$) and \cite{K07} (with a mean accuracy of
2.0\,km\,s$^{-1}$), we find that the different segments of the spectrum of a given
star yield different radial velocities, with a scatter on the level of
20\,km\,s\,$^{-1}$, indicating that there are large systematic errors in the
spectral dispersion curve deduced from the arc spectrum. The errors are traced
back to largely caused by the different light paths and focal lengths between the
arc lamp and the celestial target, as well as by the lack of sufficient number
of arc lines in the wavelength range $\lambda\lambda$5150-5800.

In order to improve the accuracy of wavelength calibration, we cross-correlate
the reduced spectra with theoretical ones of identical atmospheric parameters.
The correlation was carried out for each 250\,{\AA} spectral segment in steps
of 50\,{\AA}.  This wavelength length of    250\,{\AA} of  spectral segments was chosen based on  the following considerations: 1) Each of  spectral segment should contain  sufficient information allowing a robust  estimate of the velocity offset; 2)  A sufficient number of data points are generated ensuring a robust  polynomial fit of the offset as a function of wavelength.   The theoretical spectra were obtained by interpolating the
synthetic spectral library of \cite{Muri05} of a spectral resolution $R=2000$, similar to our observed spectra.  The resulting velocity offsets were
fitted with a  third-order  or fourth-order polynomial as a function of wavelength.  The fit was then used to
correct for errors in the wavelength calibration.  This  process is repeated twice. 
Fig.\,\ref{plot_rv} shows the  radial velocity offsets as a function of  wavelengths for the  blue and red spectra of  BD+000444.  The top panels show results before the corrections. The middle panels show the results after the first round of  corrections. Some systematic  variations of the radial velocity offsets  with wavelength remain but are acceptable considering the relatively low resolution of the spectra.  The third panels show the results after the second round of corrections. The results are quite  similar to those after the first round of  corrections, suggesting that the results have converged. 
For each blue or red spectrum,  the coefficients of the polynomial  fits are presented in the online catalogues as described in Section\,4.  
 After the corrections,
radial velocities deduced from the observed spectra with LSP3 show average offsets,
after 3$\sigma$-clipping, of only $-1.07\pm 2.02$\,km\,s$^{-1}$ and $-0.26\pm
3.82$\,km\,s$^{-1}$ for the blue and red spectra, respectively, implying a
wavelength calibration accuracy on the level of 4\,km\,s$^{-1}$(Fig.\,\ref{figurew}).

\subsection{Flux calibration}

For both spectral synthesis of stellar populations and stellar atmospheric
parameter determinations, accurate flux calibration of the SEDs of the template
stars is highly desired.
For this purpose, several effects need to be
considered and corrected for: the instrumental spectral response curve, the
wavelength-dependent flux loss for narrow-slit observation due to the
atmospheric differential refraction, the atmospheric extinction, and the
interstellar reddening. To accurately determine the instrumental spectral
response curve, each observing  night we usually observed each observing night 3 and minimum 2
spectrophotometric standard stars selected from the list available at the ESO
website.  Hot stars with few or weak spectral features were preferred.
Table~\ref{table3} lists the standards used in our observational campaign. As
described above, each target was observed with both narrow and wide slits.
Spectra obtained with the wide-slit were flux-calibrated with the response
curve derived from the standards (also observed with the wide-slit). The 
narrow-slit spectra of the targets were then scaled to match the SEDs of the 
wide-slit spectra. The atmospheric extinction was corrected for using the 
nominal extinction curve of the site.

The interstellar reddening  was corrected for using the reddening law of
\cite{1999PASP..111...63F} assuming a total-to-selective extinction ratio
$R(V)=3.1$. Values of selective extinction, i.e. colour excess $E(B-V)$, of all PASTEL stars have been derived
by Huang et al.  (2018, in preparation) using the ``star pair''  technique
of \cite{2013MNRAS.430.2188Y}.  For stars in the MILES library, the
extinction values given in the library were used.  Huang et al. (2018, in preparation) have compared values of   $E(B-V)$ derived from the ``star-pair'' technique and those provided  in MILES,  and found that the differences are mostly smaller than 0.1\,mag. The difference is a function of $E(B-V)$ in the sense that the larger the values of $E(B-V)$, the larger the differences. The standard deviation of the differences is about 0.07\,mag.  Note that for both the
purposes of spectral synthesis of stellar populations and stellar atmospheric parameter 
determinations, accurate relative, rather than absolute, flux-calibration is
sufficient.

\subsection{Notes}
Beyond 7800\,{\AA}, the spectra are affected by interference fringing caused by: (1) the different light paths between  celestial targets   and the dome flat in each night;
 (2) the delay of observation of celestial targets relative to the dome flat, which leads to  different  temperature  and thickness of the CCD  between the celestial targets  and dome flat
observations. Typically, the fringing cause fake features with strength of 5 per\,cent of the continue level, while about 230 of 1,324 red spectra are affected by
strong interference fringing whose strength are comparable to that of the absorption lines. 
The interference fringing was found hardly to removed from the spectra entirely, and it
is a defect of the current data. Stars affected by strong interference fringing are marked by 'flag-f' of value zero, while those unaffected by  'flag-f' of  value  one in the catalogues as described  in Section\,4.

The red spectra are also affected by a number of strong telluric absorption
features, notably the O$_2$ bands at $\sim 6280$ and 6870\,{\AA} and the H$_2$O
bands at $\sim 7180$, 7800 and 8200\,{\AA}. We opt not to rectify those
features in our spectra, given the  large uncertainties that might be
introduced for any correction algorithm. Stars strongly affected by  telluric absorption features are marked by 'flag-t'  of value  zero, the others by 'flat-t' of value  one.

 \begin{table*}
\centering
\caption{Spectrophotometric standard stars used in the current campaign (cf.
http://www.eso.org/sci/observing/tools/standards/spectra/stanlis.html).}
\begin{tabular}{ccccc}
\hline
Star & RA(J2000.0) & Dec.(J2000.0) &V (mag)& Spectral type\\
       &  hh:mm:ss & dd:mm:ss &    &     \\
\hline
HR153 & 00:36:58.30 & +53:53:48.9 & 3.66  & B2IV\\
HR718 & 02:28:09.54 & +08:27:36.2 & 4.28  & B9III \\
HR1544 & 04:50:36.69 &+08:54:00.7 &4.36  & A1V\\
G191-B2B & 05:05:30.62 &+52:49:54.0 & 11.78 & DA1 \\
BD+75d325 & 08:10:49.31 & +74:57:57.5 & 9.54  & O5p \\
HR3454 & 08:43:13.46 & +03:23:55.1 & 4.3  & B3V \\
Feige34 & 10:39:36.71 & +43:06:10.1 & 11.18  & DO \\
HD93521 & 10:48:23.51 & +37:34:12.8 &7.04  &  O9Vp \\
HR4468 & 11:36:40.91 & $-$09:48:08.2 & 4.7 &  B9.5V \\
Feige56 & 12:06:47.3 & +11:40:13 &11.06 &  B5p \\
Feige66 & 12:37:23.55 & +25:04:00.3 & 10.50  & sd0\\
HR4936 & 13:09:56.96 & $-$05:32:20.5 & 4.38  & A1IV \\
HR5501 & 14:45:30.25 & +00:43:02.7 & 5.68 & B9.5V\\
BD+33d2642 & 15:51:59.86 & +32:56:54.8  &10.81&B2IV\\
HR7001      & 18:36:56.33 & +38:47:01.1  & 0.00    & A0V\\
 HR7596       &19:54:44.80 & +00:16:24.6   &5.62  &A0III\\
 HR7950      & 20:47:40.55  &$-$09:29:44.7 & 3.78  & A1V\\
 BD+28d4211  & 21:51:11.07 & +28:51:51.8  &10.51  &  Op\\
 BD+25d4655   & 21:59:42.02  &+26:25:58.1  &  9.76  &  O \\
 HR8634     &  22:41:27.64  &+10:49:53.2  & 3.40  &B8V\\
 Feige110    & 23:19:58.39  &$-$05:09:55.8  &11.82  &  DOp \\
\hline 
\end{tabular}
\label{table3}
\end{table*}

\subsection{Combining blue and red spectra}

\begin{table*}
\centering
\caption{Spectral libraries used in comparison with our  LEMONY spectra.}
\begin{tabular}{ccccc}
\hline
Library & Wavelength coverage ({\AA}) & FWHM Resolution ({\AA}) & Number of stars & Comments \\
\hline
MILES & 3525-7500 & 2.5 & 985 & Long-slit spectra with good flux-calibration\\
ELODIE& 4100-6800 & 0.5 & 1347 & Echelle spectra \\
CaT & 8348-9020 & 1.5 & 706 & Long-slit spectra with good flux-calibration \\
\hline
\end{tabular}
\label{table4}
\end{table*}

\begin{figure}
\centering
\includegraphics[width=3.5in]{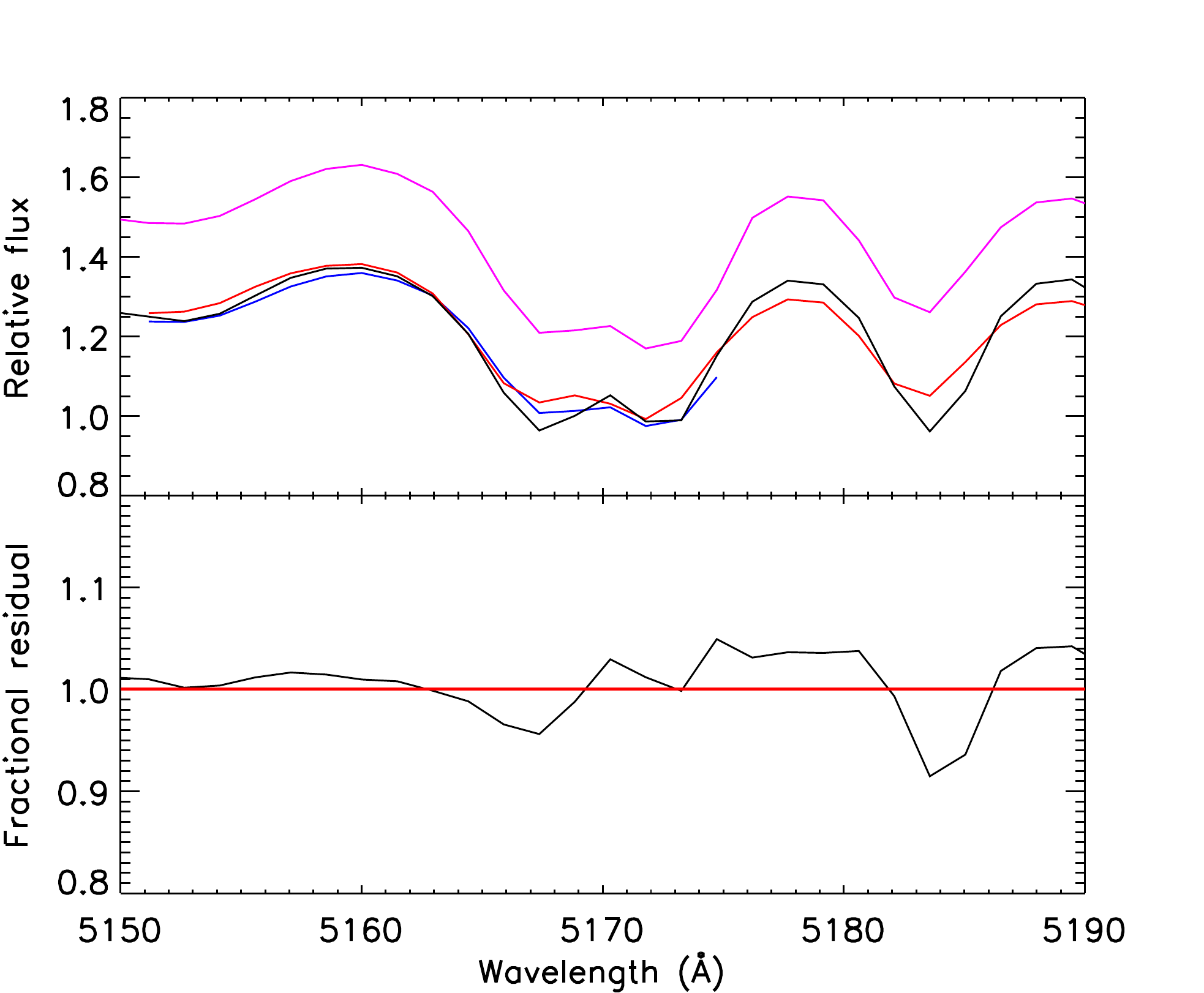}
\caption{The MILES (black line), the LEMONY blue (blue line), red (red line)  and the final combined  (scaled by a factor of 1.2 for clarity; magenta line) spectra  of  star HD054810 in the wavelength range from 5150 to 5190\,\AA \, are shown in the upper panel.  The fractional differences between the MILES and our combined spectra are  plotted in the lower panel. The horizontal red line indicates zero  difference.}
\label{figure6}
\end{figure}

Finally, the blue and red spectra of each star are combined to produce a well
flux-calibrated spectrum covering almost the whole optical wavelength range of
$\lambda\lambda$3800--9000. The blue spectra are broadened to match the resolution of the red spectra.
  Given that  blue spectra often  have a 
higher SNR  and are better calibrated than  the red ones in their overlapping region 
from 5150 to 5180\,\AA, 
the blue spectra are used for the combined spectra.  
Mg$_{1}$ and Mg$_{2}$ lines are located 
in the overlapping region, we check their profiles by 
 comparing with the MILES spectra. 
Fig.\,\ref{figure6} shows an example of  combining the red and blue spectra 
for the overlapping wavelength range  5150--5190\,\AA. The LEMONY and MILES spectra are convolved with gaussians 
to a resolution of FWHM $\sim$ 3.6\,\AA\,  in order to achieve the comparison.
The Mg$_{1}$ and Mg$_{2}$ line profiles given by the MILES spectrum and by our the final combined spectrum
 match quite  well, with fractional   differences  smaller than 10\,per\,cent. 
As discussed in Section 4.2,  the differences of Lick indices deduced are  also very small.

\section{Data access}
The spectra  are available online\footnote{link to be added once the paper has been accepted for publication} for retrieval at  \url{http://lamost973.pku.edu.cn/site/data}.  The spectra are in FITS format. Note that  spectra without correcting for the systematic errors in wavelength calibration are also included. Catalogues containing information of the template stars  are also provided, again in FITS format.  The  information includes  the  star name, coordinates, SNRs of the  blue and red spectra, spectral wavelength range, observational instrument used, observational Julian date, stellar parameters, flags of the spectra and the polynomial fit coefficients used to correcting for the systematic errors in wavelength calibration.  A Readme file is included with the online data.  A detailed description of the catalogue content is presented in Table\,\ref{table_catalogue}. 

 \begin{table*}
 \caption {Description of the catalogues.} 
 \begin{tabular}{cc}
 \hline
 Name & Description \\
\hline
name & The name of the star \\
ra & Right ascension of J2000.0 ($\circ$)\\
dec & Declination of J2000.0 ($\circ$) \\
SNR-B & Mean SNR per\,pixel of blue spectra \\
SNR-R & Mean SNR per\,pixel of red spectra \\
JD-B & Julian date at the middle time of the first exposure of the blue spectra \\
JD-R & Julian date at the middle time of the first exposure of the red spectra \\
wavelength range & The wavelength range available \\
instruments & The using telescope and spectrograph when observing the star \\
teff & The effective temperature of the star \\
logg & The surface gravity of the star \\
feh & The iron to hydrogen abundance ratio of the star \\
ebv & The adopted $E (B-V)$ \\
a1-B & The zero-order coefficient of the polynomial fits of velocity offsets as a function of  wavelength for the  blue spectra for the first time correction \\ 
a2-B & The first-order coefficient of the polynomial fitts of velocity offsets as a function of wavelength for the  blue spectra  for the first time correction\\
a3-B & The second-order coefficient of the polynomial fits of velocity offsets as a function of  wavelength for the  blue spectra for the first time correction\\
a4-B & The third-order coefficient of the polynomial fits of velocity offsets as a function of  wavelength for the  blue spectra for the first time correction\\
a5-B & The fourth-order coefficient of the polynomial fits of velocity offsets as a function of  wavelength for the  blue spectra for the first time correction\\
b1-B & The zero-order coefficient of the polynomial fits of velocity offsets as a function of  wavelength for the  blue spectra for the second time correction\\ 
b2-B & The first-order coefficient of the polynomial fits of velocity offsets as a function of  wavelength for the  blue spectra for the second time correction\\
b3-B & The second-order coefficient of the polynomial fits of velocity offsets as a function of  wavelength for the  blue spectra for the second time correction\\
b4-B & The third-order coefficient of the polynomial fits of velocity offsets as a function of  wavelength for the  blue spectra for the second time correction\\
b5-B & The fourth-order  coefficient of the polynomial fits of velocity offsets as a function of  wavelength for the  blue spectra for the second time correction\\
a1-R & The zero-order coefficient of the polynomial fits of velocity offsets as a function of  wavelength for the  red spectra for the first time correction\\
a2-R & The first-order  coefficient of the polynomial fits of velocity offsets as a function of  wavelength for the  red spectra for the first time correction\\
a3-R & The second-order  coefficient of the polynomial fits of velocity offsets as a function of  wavelength for the  red spectra for the first time correction\\
a4-R & The third-order  coefficient of the polynomial fits of velocity offsets as a function of  wavelength for the  red spectra for the first time correction\\
b1-R & The zero-order  coefficient of the polynomial fits of velocity offsets as a function of  wavelength for the  red spectra for the second time correction\\
b2-R & The first-order coefficient of the polynomial fits of velocity offsets as a function of  wavelength for the  red spectra for the second time correction\\
b3-R & The second-order coefficient of the polynomial fits of velocity offsets as a function of  wavelength for the  red spectra for the second time correction\\
b4-R & The third-order coefficient of the polynomial fits of velocity offsets as a function of  wavelength for the  red spectra for the second time correction\\
flag-s\,$^{a}$ & Mark the observational status \\
flag-f\,$^{b}$ & Mark whether the red spectrum is affected by interference  fringing \\
flag-t\,$^{c}$ & Mark whether the red spectrum is affected by telluric absorption \\
flag-v\,$^{d}$ & Mark whether the star is a variable star\\
   \hline  
  \end{tabular}
\begin{tablenotes}
\item[1] flag-s $^{a}$ mark the observational status. 0 mean that the star only has blue spectrum.  1 mean that the star only has red spectrum. 2 mean that the star has both blue and red spectra.
\item[2] flag-f $^{b}$ mark whether the red spectrum is affected by interference fringing strongly.  0 mean that the red spectrum id affected by interference fringing strongly. 1 mean that the effects of interference fringing on the red spectrum are negligible.
\item[3] flag-t $^{c}$ mark whether the red spectrum is affected by telluric absorption strongly.  0 mean that the red spectrum is affected by telluric absorption strongly. 1 mean that the effects of telluric absorption on the red spectrum are negligible. 
\item[4] flag-v\,$^{d}$ mark whether the star is a variable star. 'y' mean that the star is a variable star. 'n' mean that the star is not a variable star. 
\end{tablenotes}
 \label{table_catalogue}
 \end{table*}

\section{SPECTRAL QUALITY}

In this Section, we assess the quality of the LEMONY spectra. In order to examine the SED accuracy, we make two
comparisons: 1) Comparison of broad band colors calculated from our  spectra with those calculated from MILES spectra and the photometric measurements in the literature; 2) Direct
comparison between the LEMONY and those in the MILES and CaT \citep{1997ApJS..111..377W} libraries
 for the common targets  in the overlapping spectral regions.
To check the strengths of spectral features, we compare the Lick/IDS indices as
defined in \cite{1997ApJS..111..377W} derived from our spectra with those
deduced from the spectra in the MILES and ELODIE libraries
for the common targets. We also compare the values of a new set of the near-IR indices
(CaT$^{\star}$, CaT and PaT) defined by \cite{2001MNRAS.326..959C} deduced from
our and CaT spectra for the common objects. Table\,\ref{table4} summarize the
information of the three libraries used for the above comparisons.  The spectra from all 
the libraries are broadened to match the poorest spectral resolution (average resolution FWHM of  $\sim$3.6\,\AA\,for the LEMONY red spectra) of these
libraries, and binned to a common linear dispersion of 1.0\,\AA\,$\rm pixel^{-1}$. 

\subsection{SEDs}

\begin{figure}
\centering
\includegraphics[width=3.5in]{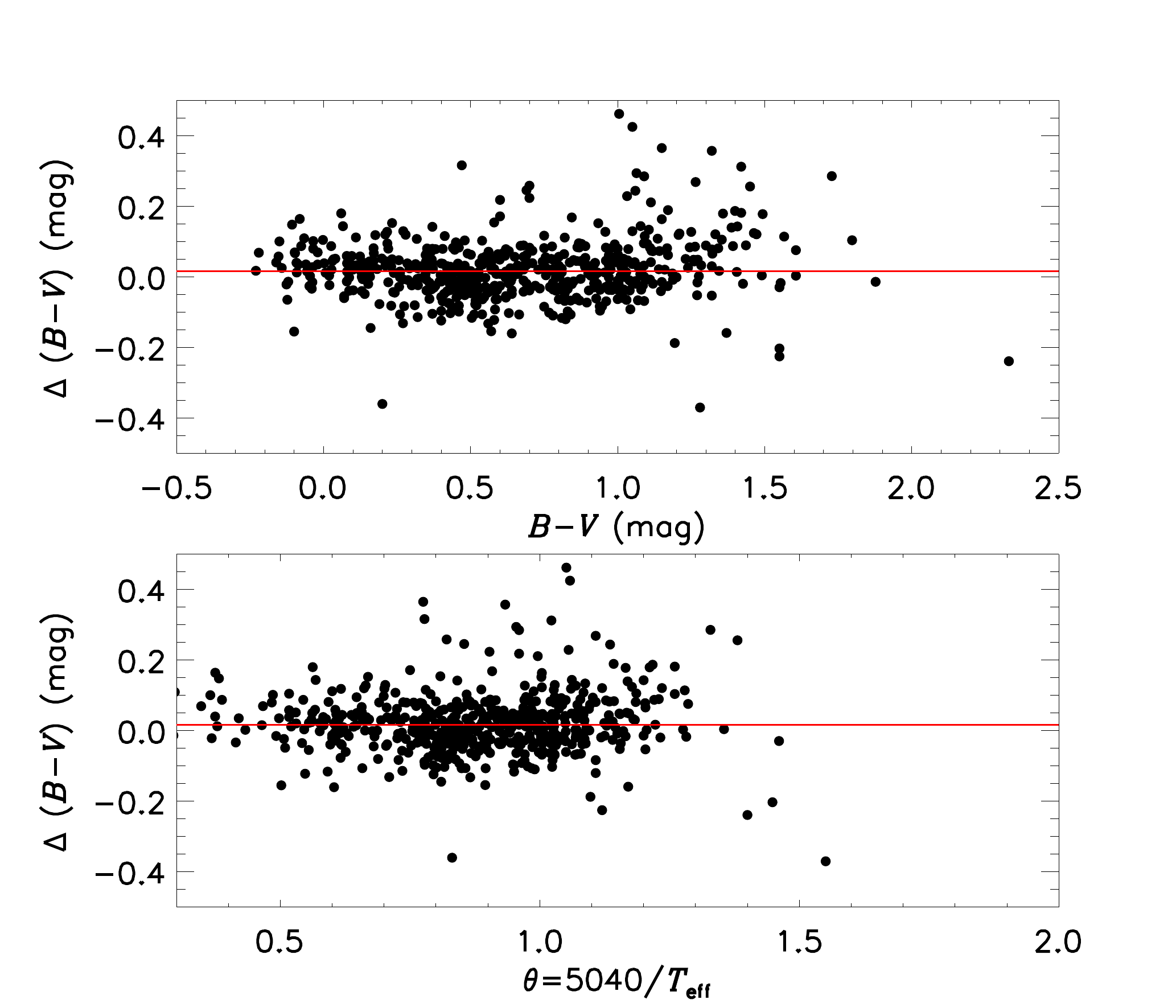}
\caption{Differences of $B-V$ colors derived from the LEMONY spectra and the photometric
values from the Lausanne database, plotted against the photometric values (top panel)
and effective temperature $T_{\rm eff}$ (bottom panel).}
\label{figure7}
\end{figure}

\begin{figure}
\centering
\includegraphics[width=3.5in]{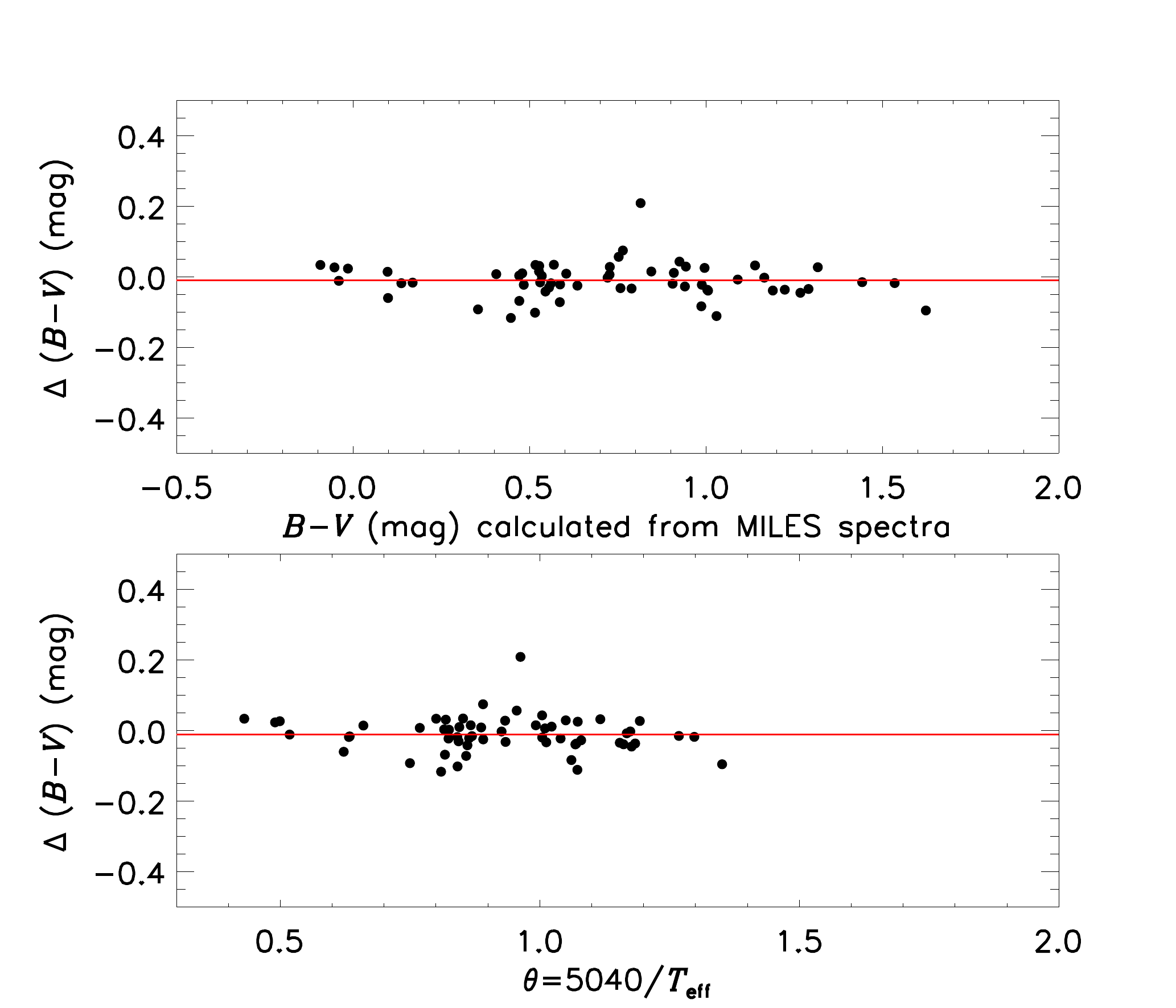}
\caption{Differences of $B-V$ colors derived from the LEMONY spectra and from MILES spectra for 60 common stars, plotted against the $B-V$ colors calculated from the MILES spectra (top panel)
and effective temperature $T_{\rm eff}$ (bottom panel).}
\label{observe_miles}
\end{figure}

From the 628 combined observed spectra (without the interstellar reddening corrections),
we calculate the values of broad band color $B-V$ using the filter transmission
curves of  \cite{1953ApJ...117..313J}. The results are compared  with values taken from the Lausanne photometric database  \citep{Mermilliod1996}. The results
are  shown in Fig.\,\ref{figure7}. The differences are essentially  smaller than 0.2\,mag. 
Some red stars (with large values of $B-V$) have differences larger than 0.3\,mag, but the relative differences remain small. 
The average difference is $0.016\pm 0.07$\,mag.  The small offset  sets an upper limit on the systematic uncertainties of
our spectral flux-calibration for the spectral range  up to $\sim$\,6000\,\AA.   The standard deviation of the difference  is of  0.07\,mag.
 For the 60 common targets, we compare  the  values of broad band colours calculated from the combined observed spectra and from the MILES spectra. The results are shown in  Fig.\,\ref{observe_miles}.   The differences are essentially  smaller than 0.1\,mag with  one star of a difference  $\sim$\,0.2\,mag. The average difference is $-0.01$\,mag, with a standard deviation of  0.043\,mag. Both the comparisons suggest an accuracy of relative flux calibration of a few  per\,cent. 

\begin{figure}
\centering
\includegraphics[width=3.5in]{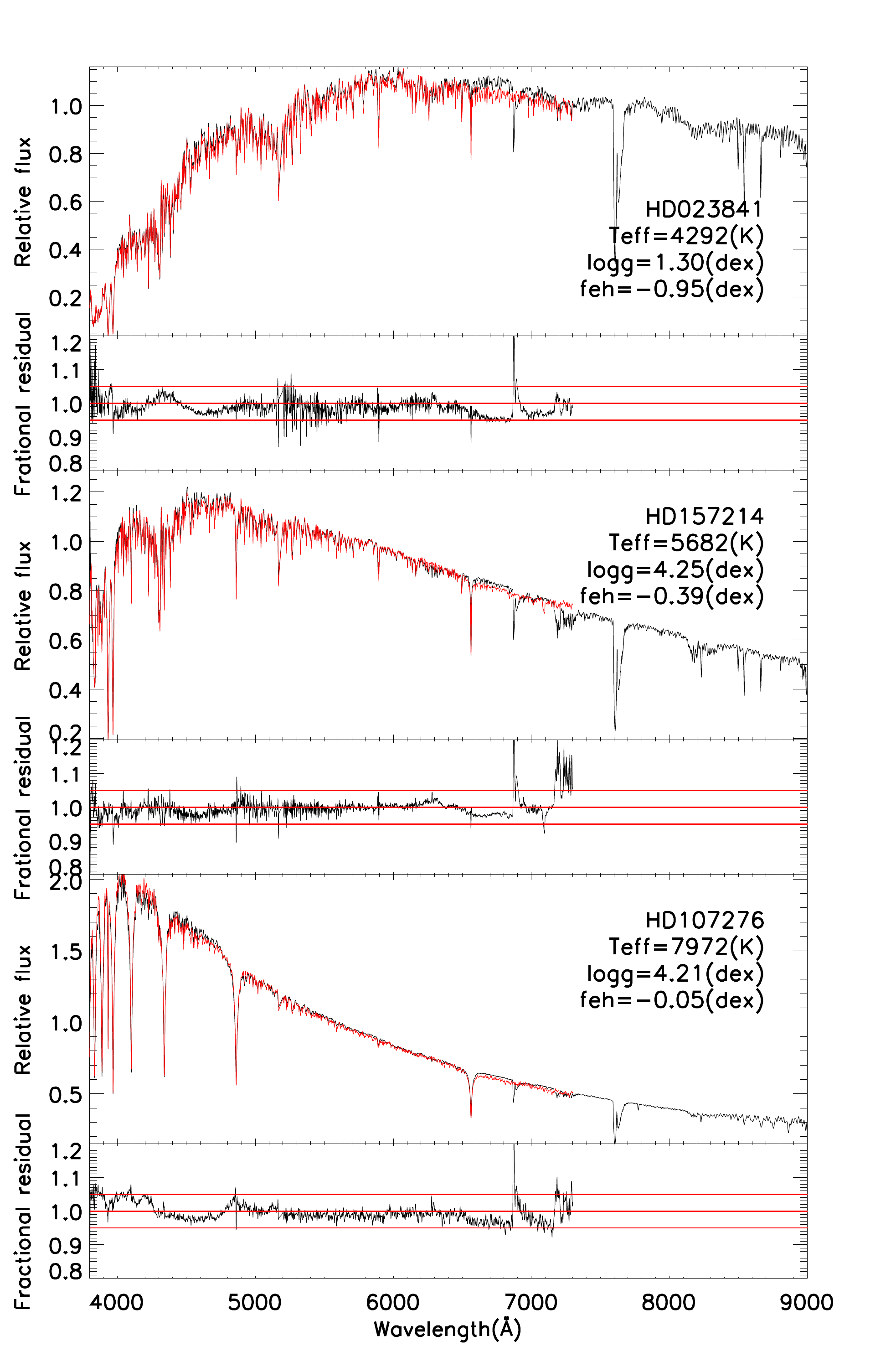}
\caption{Comparison of the LEMONY spectra (black lines) and those in the  
MILES library (red lines) for three common targets of different spectral types. The fractional
differences are also overplotted at the bottom of each panel. The horizontal
lines indicate zero and plus/minus 5 per cent differences, respectively. 
The star name and atmospheric  parameters are marked in each panel.}
\label{figure8}
\end{figure}

\begin{figure}
\centering
\includegraphics[width=3.5in]{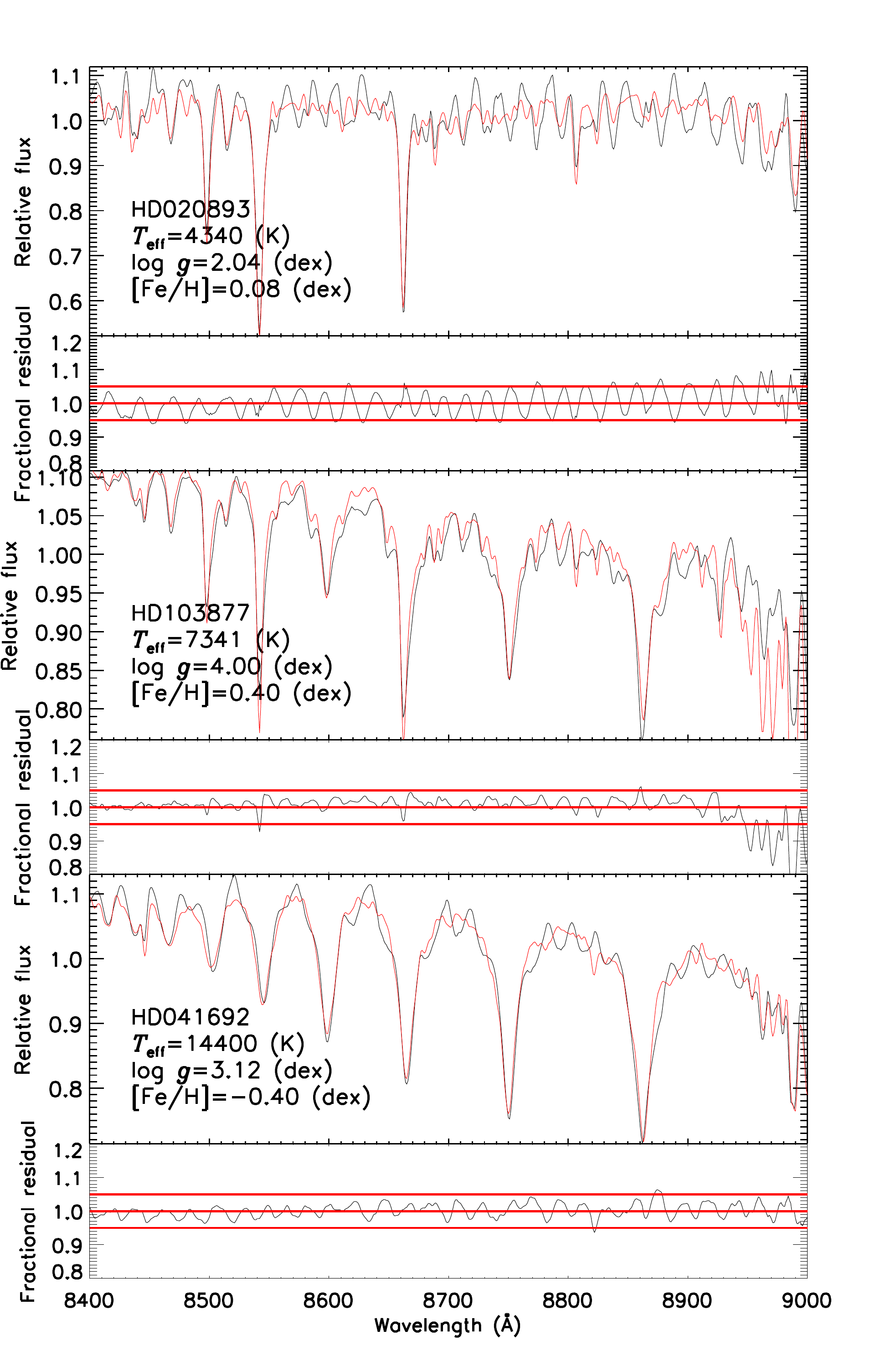}
\caption{Comparison of our spectra (black lines) and those in the CaT library
(red lines) for three  common objects of different spectral types. The fractional differences are also 
overplotted at the bottom of each panel. The horizontal lines indicate zero 
and plus/minus 5\,per\,cent differences, respectively. The star name and 
atmospheric parameters are marked in each panel.}
\label{figure9}
\end{figure}

In Fig.\,\ref{figure8} we make a direct comparison of our spectra with those from
the MILES library for three common targets.  
In all cases, the fractional differences have a
mean and a standard deviation ranging from 1 to 4 per cent.  Fig.\,\ref{figure9}
compares our spectra with those from the CaT library for three common objects
for the common wavelength range 8400--9000\,{\AA}. Again the agreement is very
good, with fractional differences of just a few per cent.  Considering the
uncertainties of  flux-calibration of the MILES and CaT spectra, the
telluric absorption features as well as the CCD interference fringing in the
far red, the comparisons shown in Fig.\,\ref{figure8} and Fig.\,\ref{figure9}
indicate that our flux-calibration is likely to be much better than 5 per cent.

\begin{figure}
\centering
\includegraphics[width=3.5in,height=3.5in]{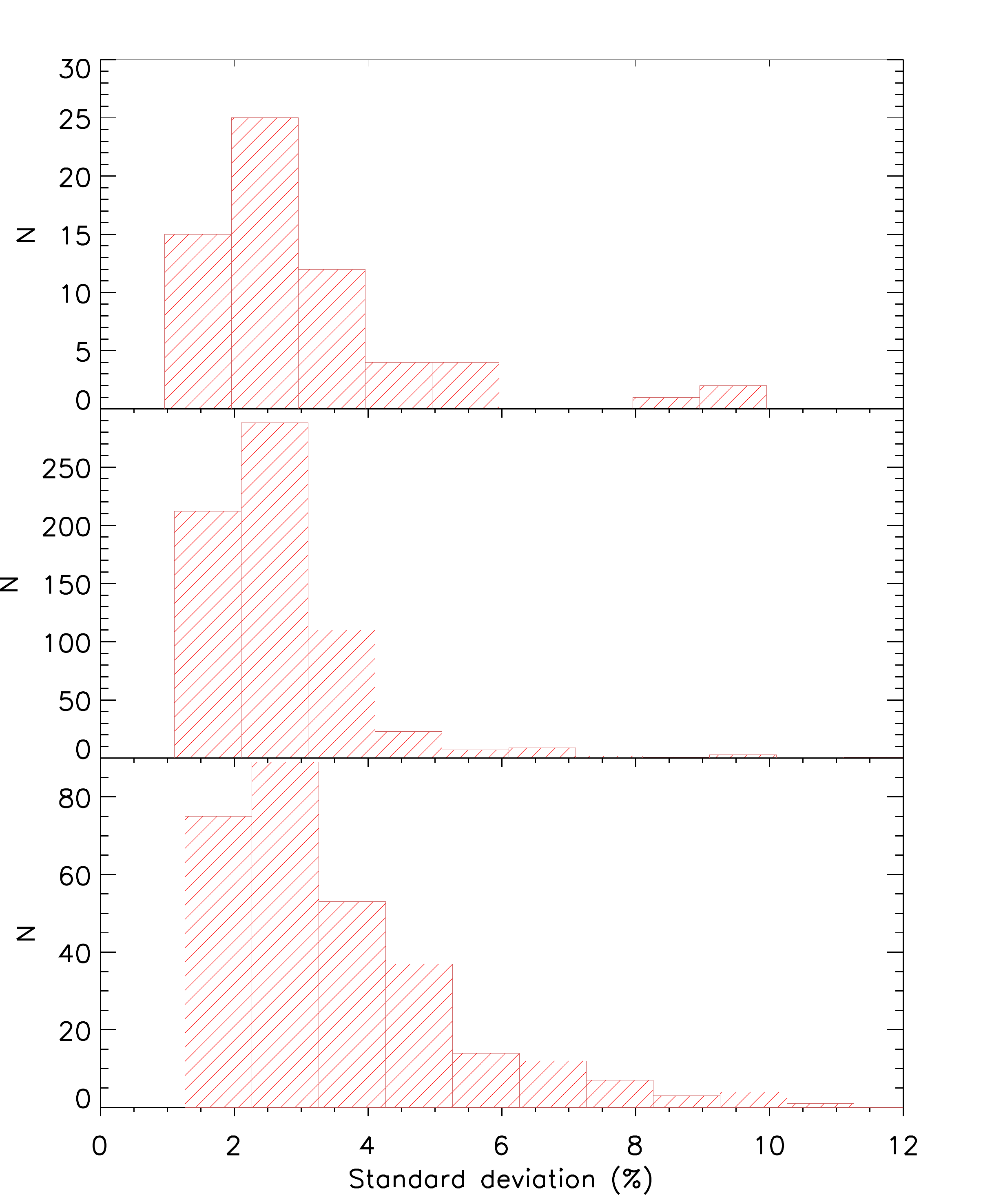}
\caption{Distributions of standard deviations of fractional differences, 
after 3$\sigma$-clipping, of our blue (top panel) and red (middle panel) spectra 
and those from the MILES library, and of our red spectra
and those from the CaT library (bottom panel).}
\label{figure10}
\end{figure}

The standard deviations  of fractional differences from three comparisons are also caculated:
1) A comparison  between our blue spectra and those   from the MILES library   for  63
 common targets for the common  wavelength range $\lambda \lambda $3800--5150;  2) A comparison  
 between our red spectra and  those   from the MILES library for 667  common targets 
 for the common wavelength range  $\lambda \lambda $5200--7400; 3) A comparison between our red spectra
and those from the CaT library for 296 common stars for the common wavelength range 
$\lambda \lambda $8400--9000.   Fig. \ref{figure10} shows the standard 
deviation distributions of the three comparisons. 
In all comparisons the fractional differences are by dividing  the LEMONY spectra and  the MILES library or the CaT library.
The resultant fractional standard deviations are essentially  all  smaller than 10\,per\,cent. 

In conclusion, the relative flux-calibration of the spectra in the LEMONY library is likely to be better 
than  5\,per\,cent. 

\subsection{Lick/IDS and the near-IR indices}

\begin{figure*}
\centering
\includegraphics[width=5.5in]{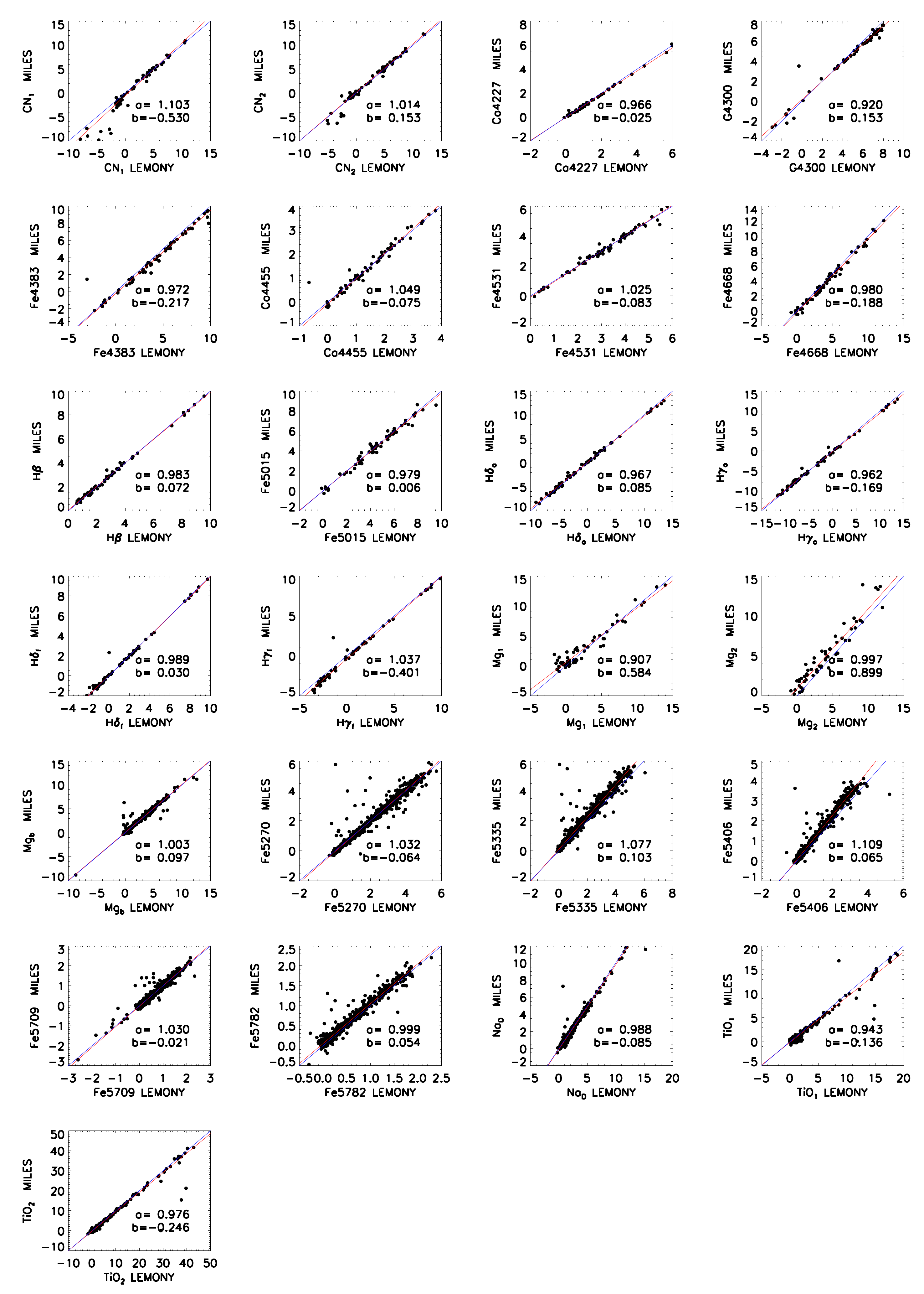}
\caption{Comparison of Lick/IDS indices derived from spectra in the 
MILES library and from the LEMONY spectra. The red  and blue line  in each panel is a linear fit to the data and a identity line, respectively. The slope ($a$) and interpret  ($b$) of the fit are marked in each panel.}
\label{figure11}
\end{figure*}

\begin{figure*}
\centering
\includegraphics[width=5.5in]{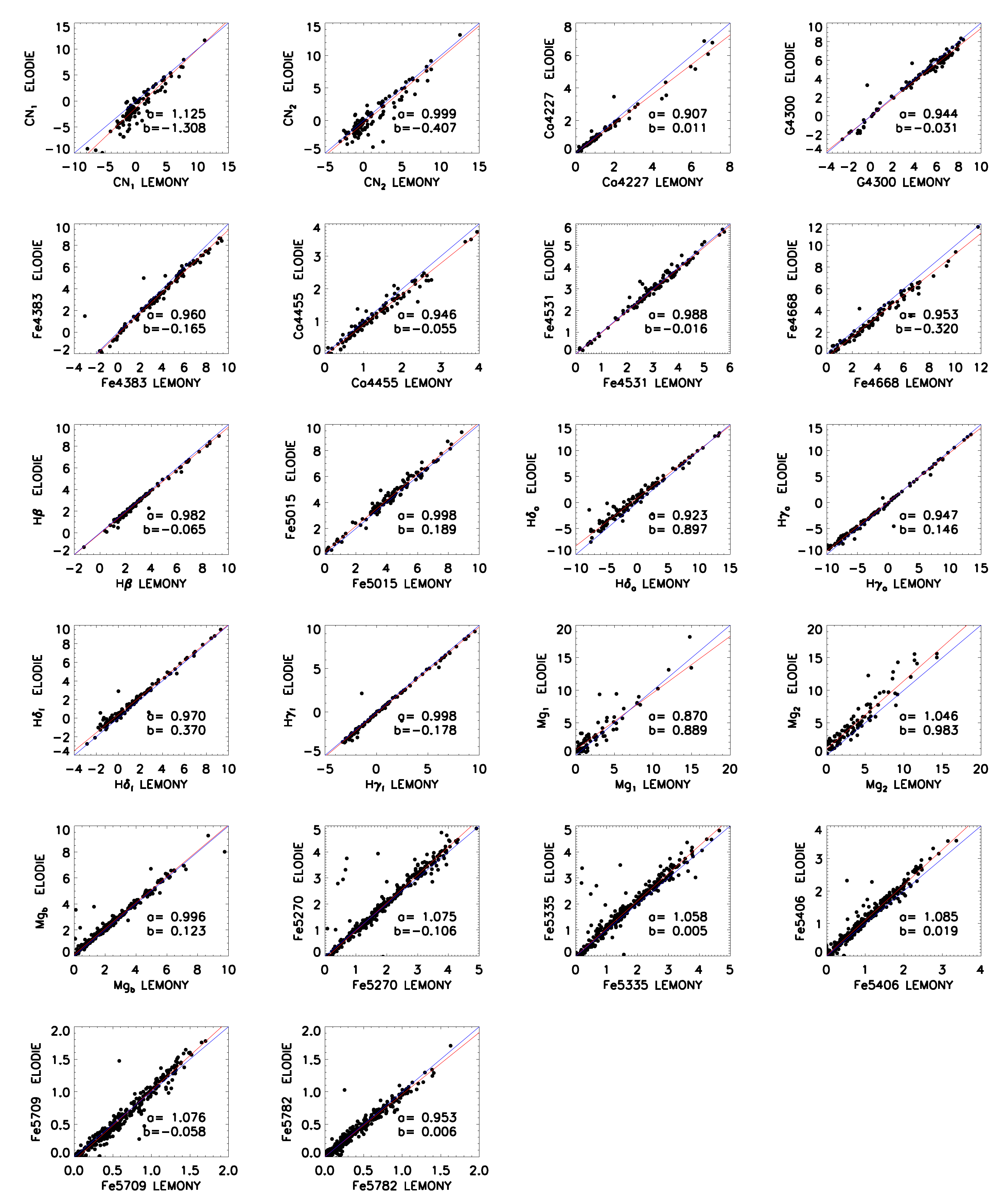}
\caption{Same as Fig.\,\ref{figure11}, but for a comparison of Lick/IDS indices derived from spectra in the ELODIE library and from  the LEMONY spectra.}
\label{figure12}
\end{figure*}

\begin{figure*}
\centering
\includegraphics[width=4.125in]{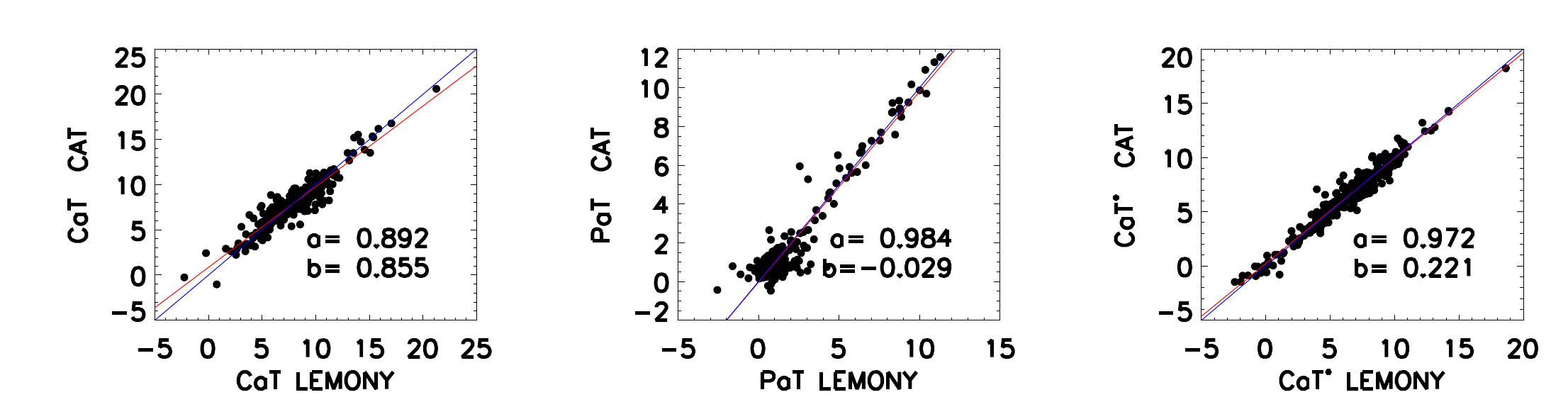}
\caption{Same as Fig.\,\ref{figure11}, but for a comparison of the near-IR indices derived from spectra in the CaT library and from the LEMONY spectra.}
\label{figure13}
\end{figure*}

The Lick/IDS indices are derived from  the LEMONY spectra and  from spectra  in the MILES 
 and ELODIE libraries for common objects.  
The results  are compared in Figs.\,\ref{figure11}\,and\,\ref{figure12}.
The near-IR indices are derived from  the LEMONY
spectra and from those in the CaT library for common stars. 
A comparison   between the near-IR indices  derived from  the LEMONY spectra 
and those deduced from spectra in the CaT library for common stars  is presented in Fig.\,\ref{figure13}.  The agreement is very good in all cases.  
There are several stars show large differences in Lick/IDS indices Fe5270, Fe5335, $\rm TiO_{1}$ and $\rm TiO_{2}$. We find that   almost all of them are variable stars including two S stars (many of them are long period variable stars).
Linear fit for the individual  indices compared, are also overplotted in the three
Figures.
The slopes ($a$), intercepts ($b$) of the individual  linear fits  are also marked in the panels.
 The  values of the slope, intercept  and RMS of residuals are also listed in Table~\ref{table5}.
 As the Table shows, all slopes have a value close 
to unity, except when comparing with the CaT library,
where the slopes are always smaller than 1, implying  some systematic differences.
  The interferometric fringing remaining  in the LEMONY spectra may be partly responsible for these differences.
The small non-zero intercepts seen in some indices also suggest the presences of  some systematic differences between the libraries. 

 \begin{table*}
 \caption {Comparisons of Lick/IDS and the near-IR indices deduced from our spectra and 
from spectra in the MILES library and in  the ELODIE and CaT libraries. For each data set, the first two
columns list the slopes and intercepts of  straight line fits to the data plotted  in Figs.\,\ref{figure11},\,\ref{figure12} and \ref{figure13}. } 
 \begin{tabular}{cccccccccc}
 \hline
 \ & \ & MILES & \  & \  & ELODIE &\  & \  & CAT &\  \\
 \ & Slope & Intercept & RMS & Slope & Intercept & RMS &Slope & Intercept & RMS  \\
 \hline
 CN$_{1}$  &     1.103  &    -0.530  &     0.527  &     1.125  &    -1.308  &     1.106  &  \  &  \  &  \ \\
CN$_{2}$  &     1.014  &     0.153  &     0.351  &     0.999  &    -0.407  &     0.981  &  \  &  \  &  \ \\
Ca4227  &     0.966  &    -0.025  &     0.066  &     0.907  &     0.011  &     0.062  &  \  &  \  &  \ \\
G4300  &     0.920  &     0.153  &     0.133  &     0.944  &    -0.031  &     0.275  &  \  &  \  &  \ \\
Fe4383  &     0.972  &    -0.217  &     0.205  &     0.960  &    -0.165  &     0.197  &  \  &  \  &  \ \\
Ca4455  &     1.049  &    -0.075  &     0.115  &     0.946  &    -0.055  &     0.109  &  \  &  \  &  \ \\
Fe4531  &     1.025  &    -0.083  &     0.143  &     0.988  &    -0.016  &     0.117  &  \  &  \  &  \ \\
Fe4668  &     0.980  &    -0.188  &     0.356  &     0.953  &    -0.320  &     0.271  &  \  &  \  &  \ \\
H$\beta$  &     0.983  &     0.072  &     0.119  &     0.982  &    -0.065  &     0.105  &  \  &  \  &  \ \\
Fe5015  &     0.979  &     0.006  &     0.308  &     0.998  &     0.189  &     0.269  &  \  &  \  &  \ \\
H$\delta_{a}$  &     0.967  &     0.085  &     0.284  &     0.923  &     0.897  &     0.596  &  \  &  \  &  \ \\
H$\gamma_{a}$  &     0.962  &    -0.169  &     0.276  &     0.947  &     0.146  &     0.254  &  \  &  \  &  \ \\
H$\delta_{f}$  &     0.989  &     0.030  &     0.109  &     0.970  &     0.370  &     0.223  &  \  &  \  &  \ \\
H$\gamma_{f}$  &     1.037  &    -0.401  &     0.183  &     0.998  &    -0.178  &     0.110  &  \  &  \  &  \ \\
Mg$_{1}$  &     0.907  &     0.584  &     0.899  &     0.870  &     0.889  &     0.901  &  \  &  \  &  \ \\
Mg$_{2}$  &     0.997  &     0.899  &     0.917  &     1.046  &     0.983  &     1.002  &  \  &  \  &  \ \\
Mg$_{b}$  &     1.003  &     0.097  &     0.145  &     0.996  &     0.123  &     0.128  &  \  &  \  &  \ \\
Fe5270  &     1.032  &    -0.064  &     0.146  &     1.075  &    -0.106  &     0.118  &  \  &  \  &  \ \\
Fe5335  &     1.077  &     0.103  &     0.171  &     1.058  &     0.005  &     0.142  &  \  &  \  &  \ \\
Fe5406  &     1.109  &     0.065  &     0.105  &     1.085  &     0.019  &     0.099  &  \  &  \  &  \ \\
Fe5709  &     1.030  &    -0.021  &     0.087  &     1.076  &    -0.058  &     0.073  &  \  &  \  &  \ \\
Fe5782  &     0.999  &     0.054  &     0.074  &     0.953  &     0.006  &     0.055  &  \  &  \  &  \ \\
Na5849  &     0.988  &    -0.085  &     0.180  &  \  &  \  &  \  &  \  &  \  &  \ \\
TiO$_{1}$  &     0.943  &    -0.136  &     0.327  &  \  &  \  &  \  &  \  &  \  &  \ \\
TiO$_{2}$  &     0.976  &    -0.246  &     0.371  &  \  &  \  &  \  &  \  &  \  &  \ \\
CaT  &  \  &  \  &  \  &  \  &  \  &  \  &     0.892  &     0.855  &     0.867 \\
PaT  &  \  &  \  &  \  &  \  &  \  &  \  &     0.984  &    -0.029  &     0.568 \\
CaT$^{*}$  &  \  &  \  &  \  &  \  &  \  &  \  &     0.972  &     0.221  &     0.648 \\

 \hline  
  \end{tabular}
 \label{table5}
 \end{table*}

\section{Improvements of LSP3 using the LEMONY library}
The LEMONY spectra will  be added to the empirical template spectral library that LSP3  uses to 
estimate stellar atmospheric  parameters from the LAMOST spectra by template matching. 
We expect this  will  improve LSP3 in at least  two aspects.  Firstly, the new spectra will expand the coverage and
 improve the homogeneity of the distribution of the  MILES 
template stars in the parameter space, reducing the  systematic errors in the derived stellar atmospheric  parameters.
Secondly, the 1,324 LEMONY red  spectra can be used as templates to estimate  stellar atmospheric  parameters 
from the  LAMOST red-arm  spectra.  
There are  $\sim$\,30 per\,cent LAMOST spectra  having a SNR  better than 10 in the red but much worse  in blue. For these stars, their stellar atmospheric 
parameters   are  expected to be  determined  using the  LAMOST red-arm spectra. 
 To test this, we have carried out  a simple test to validate  the  feasibility of deriving stellar  atmospheric  parameters using the LAMOST red-arm spectra by template matching with our  red spectra. 

\begin{figure*}
\centering
\includegraphics[width=7.0in]{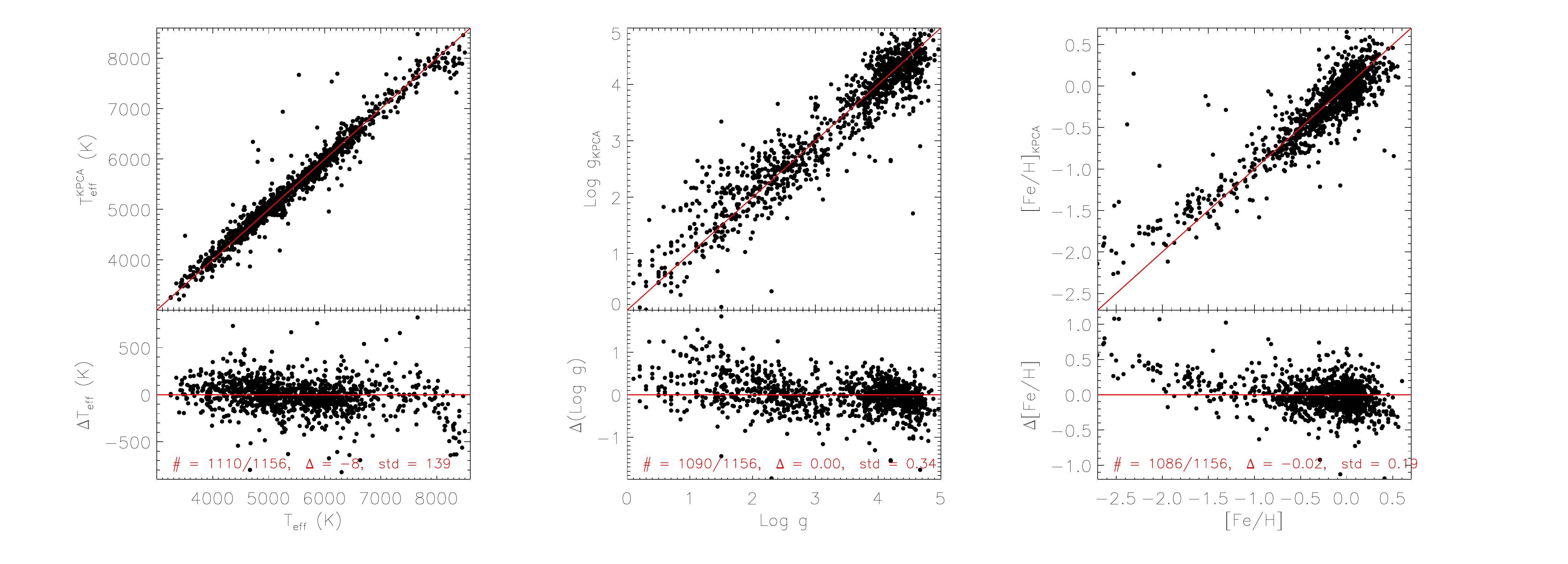}
\caption{Residuals of the KPCA fits.  The upper panels show direct comparisons of parameters given by the fits  and those from the PASTEL catalogue, while the lower panels show the  residuals.
 The number of stars after and before 3-$\sigma$ clipping, the offset  and  dispersion of the residuals are marked in the lower  panel for each compared parameter.  The red lines in the top and bottom panels represent the identity lines and zero parameters' differences, respectively.}
\label{residual}

\end{figure*}

We estimate the atmospheric  stellar parameters using spectral segments of  wavelength range 
 $\lambda\lambda$5800--9000  with  a machine learning method based on Kernel-based  principle component analysis (KPCA),
as described in  \cite{Xiang2017}.     We select stars of  3,000\,$\leq T_{\rm eff} \leq 8,500$\,K, 
[Fe/H] $\geq -3.5$\,dex and $\rm log \it\, g \rm \geq 0$\,dex   in the LEMONY library as the training set, which contains 1,156 stars in total. 
We adopt 100 principle components (PCs) in fitting the atmospheric parameters with a multi-dimensional linear function.
The residuals of the fits are shown in  Fig.\,\ref{residual}. The Figure shows that  the residuals  of atmospheric  parameters of 95\,per\,cent of the total 1,156 stars are within three standard deviations from a mean.
 The residuals  have a dispersion of 139\,K, 0.34\,dex and 0.19\,dex for $T_{\rm eff}$, 
log\,$g$ and [Fe/H] after 3-$\sigma$ clipping, respectively.

\begin{figure*}
\centering
\includegraphics[width=7.0in]{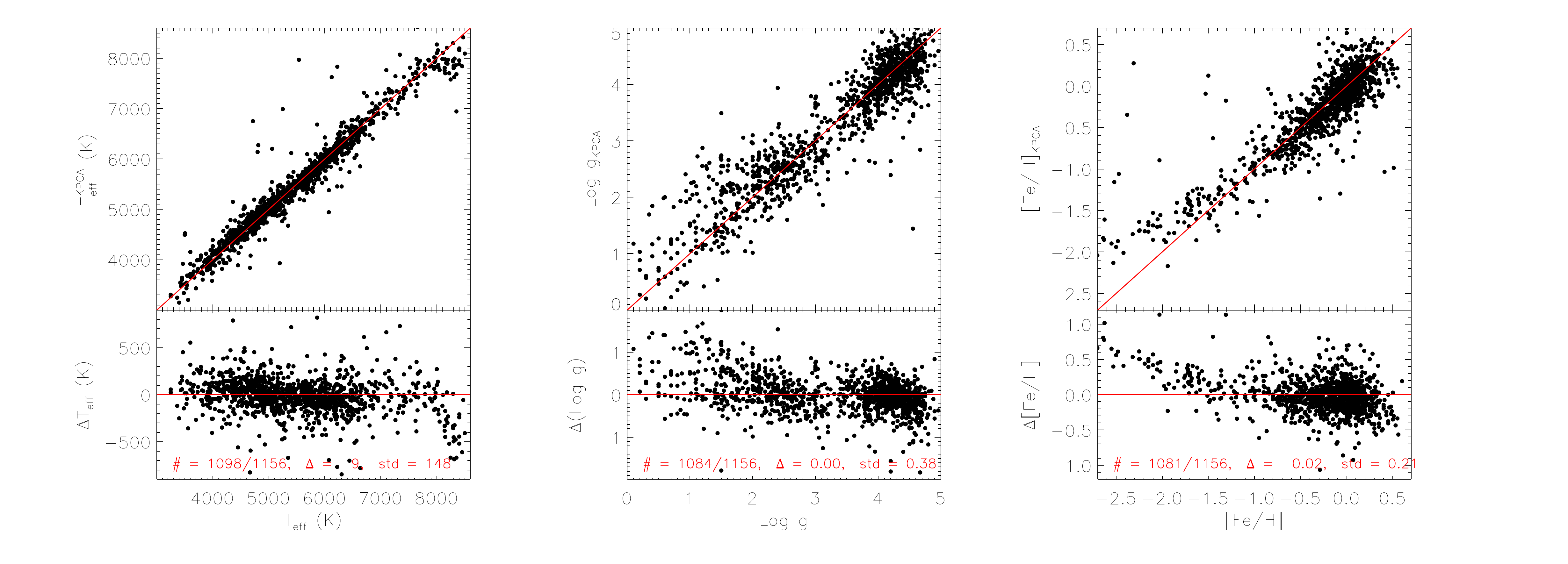}
\caption{Same as Fig.\,\ref{residual}, but for  comparisons of atmospheric  parameters deduced with the KPCA method (adopting the  leave-one-out approach) and those from the PASTEL catalogue. The red lines in the top and bottom panels represent the identity lines and zero parameter differences,  respectively.}
\label{residual_1}

\end{figure*}

 In order to  examine  whether the small dispersions of residuals are caused by over-fitting, we  have carried  out an analysis with  the so-called 
 leave-one-out approach.  
We select one star as the test sample, the other remaining 1,155 stars as the training set, and then  estimate  the atmospheric   parameters of the test star. The exercise is repeated 1,156 times, each time with a different test star.  
  Fig.\,\ref{residual_1} shows a comparison of the atmospheric  parameters thus deduced 
 with the KPCA method and  those from  PASTEL catalogue.  Again the Figure shows that  the  residuals of atmospheric  parameters of   95\,per\,cent  of the total 1,156 stars are within three standard deviations from a mean. The differences in $T_{\rm eff}$,  log\,$g$ and [Fe/H]
 are $-9\pm148$\,K, $0.00\pm0.37$\,dex and $-0.02\pm0.21$\,dex  after 3-$\sigma$ clipping, respectively.    
 The dispersions are thus comparable to those of the fitting residuals of the training set shown in Fig.\,\ref{residual}. The exercise eliminates the possibility that we have  significantly overfitted of the data.

\begin{figure*}
\centering
\includegraphics[width=7.0in]{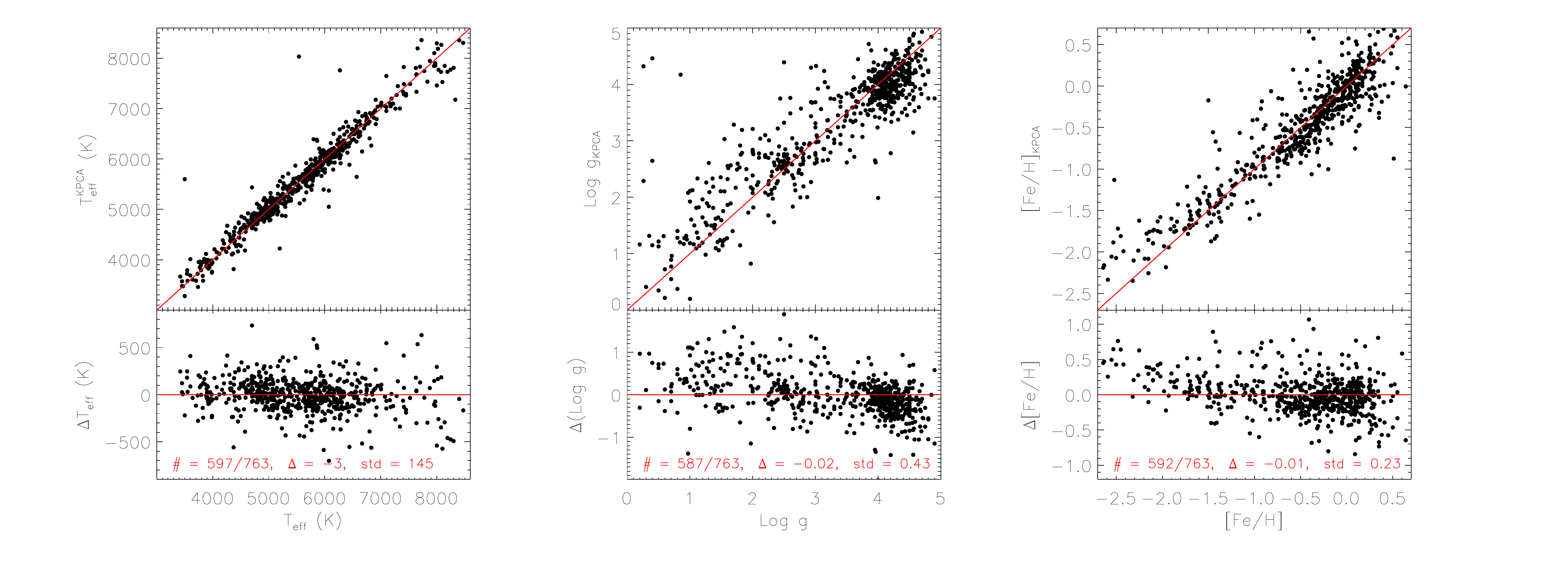}
\caption{Same as Fig.\,\ref{residual_1}, but for  comparisons of atmospheric  parameters deduced using the MILES spectra as templates.}
\label{residual_miles}

\end{figure*}

 We have also estimated the atmospheric parameters with    the leave-one-out approach  using the spectral segments of wavelength range $\lambda\lambda$5800--7400 of the 763 MILES spectra.  Fig.\,\ref{residual_miles} shows a comparison of the atmospheric  parameters deduced 
 with the KPCA method and  those from the  MILES library.  The Figure shows that   the  residuals of atmospheric  parameters of only 77\,per\,cent of the total 763 stars are within three standard deviations from a mean.  The differences in $T_{\rm eff}$,  log\,$g$ and [Fe/H]
 are $-3\pm145$\,K, $-0.02\pm0.43$\,dex and $-0.01\pm0.23$\,dex after 3-$\sigma$ clipping, respectively.  The dispersion of differences in  log\,$g$   is much larger than that derived using the  LEMONY red spectra  as the templates. 

In conclusion, when deriving the stellar atmospheric parameters using the LAMOST red-arm spectra, template matching with the LEMONY red spectra yields much results than template matching with the MILES spectra.  On the other hand,  more tests of the method are needed before applying it  to the  whole LAMOST data set, considering most of the LAMOST spectra have much lower SNRs than those of  the template stars.  

\section{SUMMARY}
In this work we built a new stellar spectral library, LEMONY, based on the observations using the OMR long-slit spectrograph mounted on the
NAOC 2.16 m telescope and the YFOSC long-slit spectrograph mounted on the YNAO 2.4 m telescope. The coverage in the parameter space is originally based on MILES, but expanded through selecting targets from PASTEL catalogue in order to improve the coverage and homogeneity of the distribution of the MILES template stars.  The wavelength coverage of the template spectra is also extended to the far red beyond the Ca~{\sc {ii}} triplet. 


Hitherto, 822 OMR (blue) and 1,324 YFOSC (red) spectra, 
covering respectively wavelength ranges $\lambda\lambda$3800--5180 and $\lambda\lambda$5150--9000, 
have been observed and reduced.  The spectra has a FWHM resolution
of about 3.3\,{\AA}, and a mean SNR higher than 100 per pixel for essentically all of
them. 
 An accuracy of $\sim 0.3$\,{\AA} and $\sim 5$\,per\,cent have been achieved for the wavelength- and flux-calibration.
   The wavelength-calibration is further improved 
to an accuracy of $\sim\,4$\,km\,s\,$^{-1}$ after corrected for the systematic errors in the spectral
dispersion relations derived from the arc spectra.  
Comparison of broad band ($B-V$) colours calculated from the LEMONY spectra with those calculated from the MILES spectra and the photometric
 measurements from the Lausanne photometric database, and   comparison between  the LEMONY
 spectra with those from the MILES and CaT libraries for the
 common stars, all suggest  that a flux-calibration accuracy of the LEMONY spectra  of  $\sim$\,5\,per\,cent. 
The Lick/IDS and the  near-IR indices derived from the LEMONY spectra are also consistent 
with those derived from the spectra in the MILES, ELODIE and CaT  libraries. 
Currently, the LEMONY library contains 822 blue and 1,324 red spectra. Together with the MILES spectra, one now has  1,731,  1,542, 1,324 and 1,273  stars with high quality spectra covering 
respectively wavelength ranges $\lambda\lambda$3800--5180, $\lambda\lambda$3800--7500, $\lambda\lambda$5150--9000 
and $\lambda\lambda$ 3800--9000.   Compared with the MILES library, the coverage and homogeneity of the distribution of the template stars in the LEMONY library in the   parameter space   are much improved. In addition, a significant fraction of the stars have red spectra extending in wavelength beyond the Ca~{\sc ii} triplet.


The LEMONY library is expected to reduce the  systematic 
errors of  atmospheric  parameters deduced with LSP3. 
 The 1,324 LEMONY red spectra will be used as template spectra to estimate atmospheric  parameters from the LAMOST red-arm spectra,
 for stars either intrinsically red 
(i.e. of late spectral types) or heavily reddened by the interstellar dust grains, increasing  the number of stars surveyed by LAMOST with atmospheric
 parameters determined by  $\sim$ 30  per\,cent. 
The LEMONY library should  also  be useful for stellar population syntheses  of galaxies and clusters  in  a wide wavelength coverage.
Of course, it should also benefit  other studies, such as the spectral classification of stars, tests of the stellar atmospheric models,etc.  
\section*{ACKNOWLEDGEMENTS}

We appreciate the helpful comments from the anonymous referee.This work was supported by National Key Basic Research Program of China
2014CB845700 and by the National Natural Science Foundation of China U1531244
and 11473001. We acknowledge the support of the staff of the YNAO 2.4\,m
telescope. Funding for the telescope has been provided by Chinese Academy of
Sciences and the People's Government of Yunnan Province. This work was
partially Supported by the Open Project Program of the Key Laboratory of
Optical Astronomy, National Astronomical Observatories, Chinese Academy of
Sciences. 

\bibliographystyle{mn2e}

\bibliography{1}

\end{document}